\documentclass{svproc}
\usepackage{url}
\usepackage{graphicx}
\usepackage{subcaption}
\usepackage{tabularx}

\usepackage[numbers,sort&compress]{natbib}
\begin{document}
	\mainmatter              
	\title{A step towards treatment planning for microbeam radiation therapy: fast peak and valley dose predictions with 3D U-Nets}
	\titlerunning{Microbeam dose predictions with 3D U-Nets}  
	%
	\author{Florian Mentzel\inst{1} \and Micah Barnes\inst{2,3} \and Kevin Kr\"oninger\inst{1} \and Michael Lerch\inst{2,4} \and Olaf Nackenhorst\inst{1} \and Jason Paino\inst{2,4} \and Anatoly Rosenfeld\inst{2,4} \and Ayu Saraswati\inst{5} \and Ah Chung Tsoi\inst{5} \and Jens Weingarten\inst{1} \and Markus Hagenbuchner\inst{5} \and Susanna Guatelli\inst{2,4}}

	\authorrunning{Florian Mentzel et al.} 
	%
	%
	\institute{
		Department of Physics, TU Dortmund University, Dortmund, NRW, Germany
		\and
		Centre for Medical Radiation Physics, University of Wollongong, Wollongong, New South Wales, Australia
		\and 
		Australian Synchrotron (ANSTO), Clayton, Victoria, Australia
		\and
		Illawarra Health and Medical Research Institute, University of Wollongong,Wollongong, New South Wales, Australia
		\and 
		School of Computing and Information Technology, University of Wollongong,Wollongong, New South Wales, Australia
	}
	
	\maketitle              
	
	\begin{abstract}
		Fast and accurate dose predictions are one of the bottlenecks in treatment planning for microbeam radiation therapy (MRT). In this paper, we propose a machine learning (ML) model based on a 3D U-Net. Our approach predicts separately the large doses of the narrow high intensity synchrotron microbeams and the lower valley doses between them. For this purpose, a concept of macro peak doses and macro valley doses is introduced, describing the respective doses not on a microscopic level but as macroscopic quantities in larger voxels. The ML model is trained to mimic full Monte Carlo (MC) data. Complex physical effects such as polarization are therefore automatically taking into account by the model.
		The macro dose distribution approach described in this study allows for superimposing single microbeam predictions to a beam array field making it an interesting candidate for treatment planning. It is shown that the proposed approach can overcome a main obstacle with microbeam dose predictions by predicting a full microbeam irradiation field in less than a minute while maintaining reasonable accuracy.
	\end{abstract}
	\section{Introduction}
	Microbeam radiation therapy (MRT) is a novel and currently pre-clinical radiotherapy treatment based on planar arrays of high intensity sub-millimetre synchrotron gamma rays \cite{Slatkin1995, Bartzsch2020, Fukunaga2021}. The narrow beams cut with a multi-slit collimator from the broad beam, often only 50$\,\mu$m wide each, result in high doses only being applied in the volumes passed by the microbeams (\textit{peak dose}), while in the space between the microbeams a significantly lower dose is deposited (\textit{valley dose}). This was observed to lead to good healthy tissue sparing and tumour control and is therefore a promising candidate for treatment of a variety of tumours which have a weak clinical outcome with currently available treatments \cite{Brauer-Krisch2015,Grotzer2015}. Ongoing research, e.g. conducted at the Australian Synchrotron (e.g. \cite{Dipuglia2019, Engels2020}) and the European Synchrotron Radiation Facility (ESRF, France, a recent overview can be found in \cite{Bartzsch2020}), has provided important steps for MRT to move towards first clinical trials. 
	\\
	Presently, accurate treatment planning for MRT is only possible using full Monte Carlo (MC) simulations which take up to 250$\,$h CPUh (hours per CPU used). A faster algorithm involving only a first-order MC simulation and applying an additional kernel method to account for the dose depositions from secondary electrons is available but still requires up to 30 minutes per configuration \cite{Donzelli2018}.
	\\
	In a recent publication it was shown that a deep learning model adapted from the so-called 3D U-Net \cite{Cicek2016} can be used to predict the dose distribution of a rectangular synchrotron broad beam of 8$\,$mm width and height in a simplified head phantom fast and accurately \cite{Mentzel2022}. In this study we extend the research to the prediction of dose distributions resulting from synchrotron microbeams. For this purpose, we propose an MRT dose prediction approach based on the superposition of individual predictions of single microbeams to a whole microbeam array field. This results in the ability to build complex irradiation fields flexibly from simple building blocks. 
	\begin{figure}[t]
		\centering
		\begin{subfigure}[t]{0.45\textwidth}
			\includegraphics[width=\linewidth]{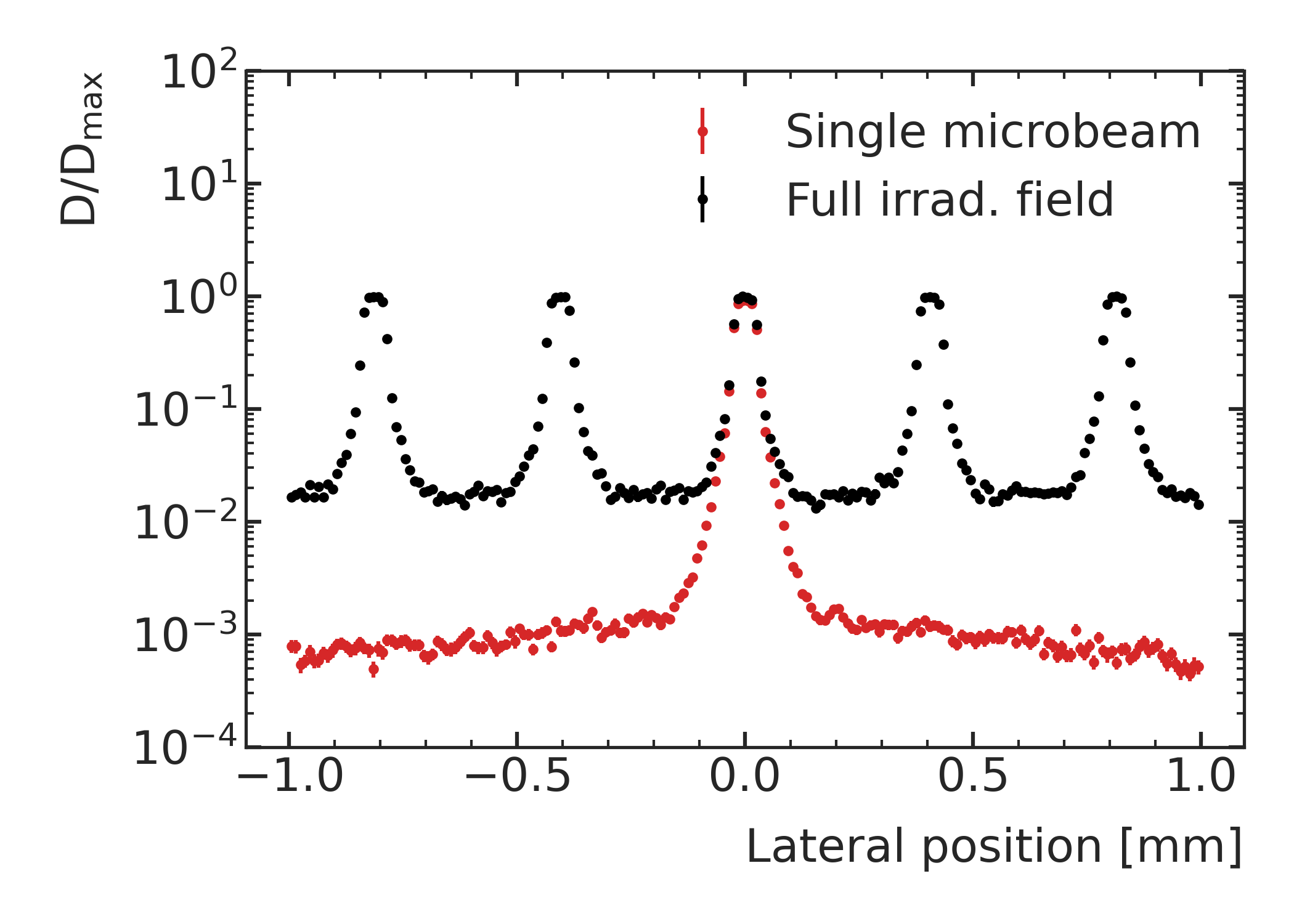}
			\caption{}
			\label{one_vs_many_microbeams:a}
		\end{subfigure}
		\begin{subfigure}[t]{0.45\textwidth}
			\includegraphics[width=\linewidth]{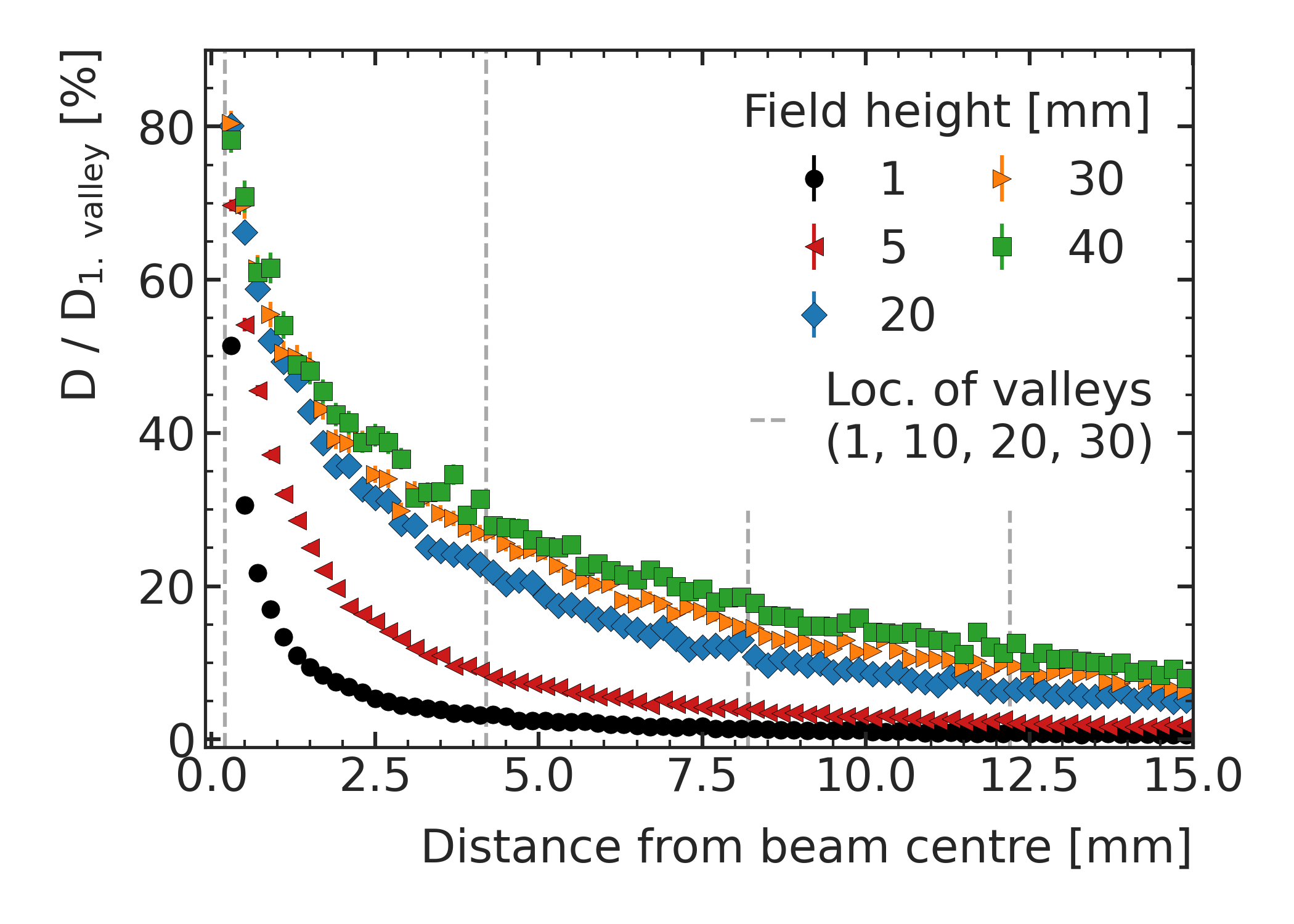}
			\caption{}
			\label{one_vs_many_microbeams:b}
		\end{subfigure}
		\caption{(a): Dose profile of a single microbeam (red) and a microbeam array (black). (b): Contribution of a microbeam located at $0\,$mm to the valleys further away relative to the contribution to its directly neighbouring valley.}
		\label{one_vs_many_microbeams}
	\end{figure}	
	Accurately predicting the dose deposited by a microbeam requires high spatial resolution, resulting in a need for small dose scoring voxels. Irradiation fields are often significantly larger ($\approx 1-20\,$mm) than one microbeam is wide. This large aspect ratio leads to a high importance of secondary electrons even far away from the actual microbeam. Figure (\ref{one_vs_many_microbeams:a}) highlights the vast difference in valley doses resulting from one microbeam compared to the full beam array. At the location of the first valley, directly next to the central microbeam peak, the valley doses differ by an order of magnitude which indicates that not only the next neighbouring microbeams can be taken into account for valley dose calculation. This issue is further investigated in Figure (\ref{one_vs_many_microbeams:b}) which shows the relative contribution of a microbeam to valleys many microbeams away. Given a large field size, a microbeam can for example contribute more than 10\% of the dose it deposits in it's adjacent valley ($\approx200\,\mu$m from the beam centre) in the valley which is located 30 microbeams (or 12.5$\,$mm) far away. 	
	As a solution to this, we propose a dose scoring method resulting in the peak and valley doses being treated as macroscopic quantities assigned to larger voxels of e.g. 0.5$\,$mm or 1$\,$mm edge length (depending on the treatment requirements) which we will refer to as \textit{macro voxels}. We will show how the resulting macro voxel data, i.e. the \textit{macro peak dose} and \textit{macro valley dose}, are suitable for superposition of dose distributions from single microbeams making them suited as training data for an ML model for microbeam dose predictions.
	\\
	The remainder of this paper is structured as follows: Section (\ref{section:methods}) presents the MC simulations which were adopted to produce the data shown in this paper. Subsequently, the novel macro voxel approach is introduced and explained in detail. Finally, the ML model is presented in Section (\ref{sectino_MLmodel}) which is trained to predict the macro peak and valley dose distributions in a simple water phantom. Section (\ref{section:results}) first presents results that support the assumption that microbeam array irradiations can be modelled as superposition of single microbeams. Thereafter, we show the results of this proof of concept ML training on the novel macro voxel data. Results and some limitations are discussed in Section (\ref{section:discussion}). A summary and an outlook is presented in Section (\ref{section:conclusion}).
	
	\section{Methods\label{section:methods}}
	This Section presents the MC simulations used to generate data samples, introduces the \textit{macro voxel} approach to data taking and presents the 3D U-Net architecture that is trained to predict the microbeam dose distributions previously simulated with the MC code.
	
	\subsection{Monte Carlo Simulations}
	All MC simulations were performed using Geant4 10.6p01 \cite{Agostinelli2003, Allison2016}. The physics list used comprises the option 4 of the electromagnetic physics constructor which is extended for polarized physics processes using the \textit{Livermore Polarized Physics} processes \cite{Geant4PhysicsManual}. 
	\\
	The microbeam particle source used in the simulations is obtained as phase space file (PSF) from a previously published and experimentally validated Geant4 simulation \cite{Dipuglia2019} of the full Imaging and Medical Beamline (IMBL) at the Australian Synchrotron, ANSTO \cite{Stevenson2017}. The microbeams have a nominal width of $w=40\,\mu$m each and a height of $h=1\,$mm. The individual microbeams are spaced at a distance of $411\,\mu$m when entering the phantom. The photons comprising the microbeams have a mean energy of approximately 80$\,$keV, the energy distribution is shown in Figure (\ref{microbeams_energy}).
	
	\subsubsection{Microbeam Array Simulation}
\begin{figure}[t]
		\centering
		\begin{subfigure}[t]{0.35\textwidth}
			\includegraphics[width=\linewidth]{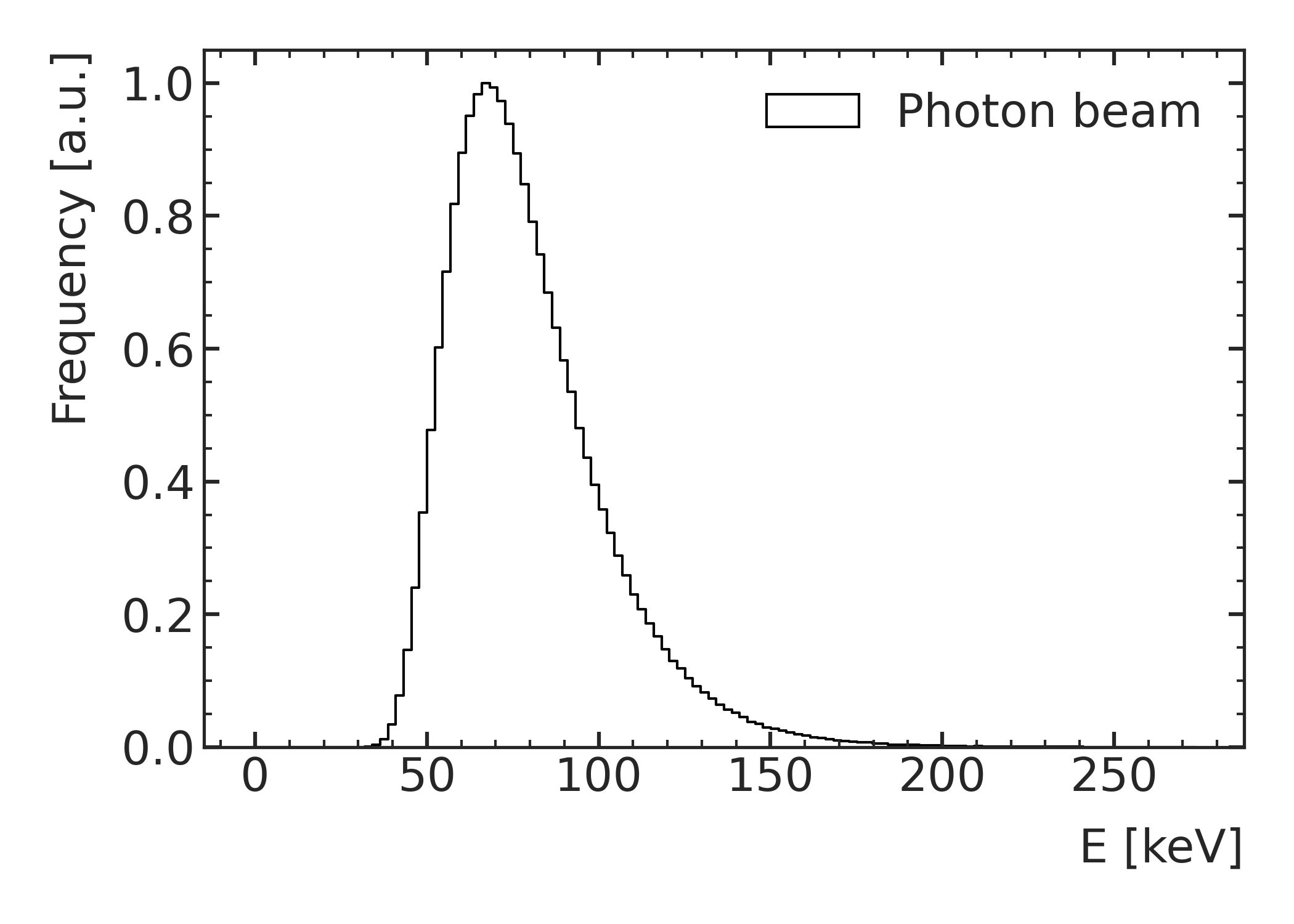}
			\caption{}
			\label{microbeams_energy}
		\end{subfigure}
		\begin{subfigure}[t]{0.6\textwidth}
			\includegraphics[width=\linewidth]{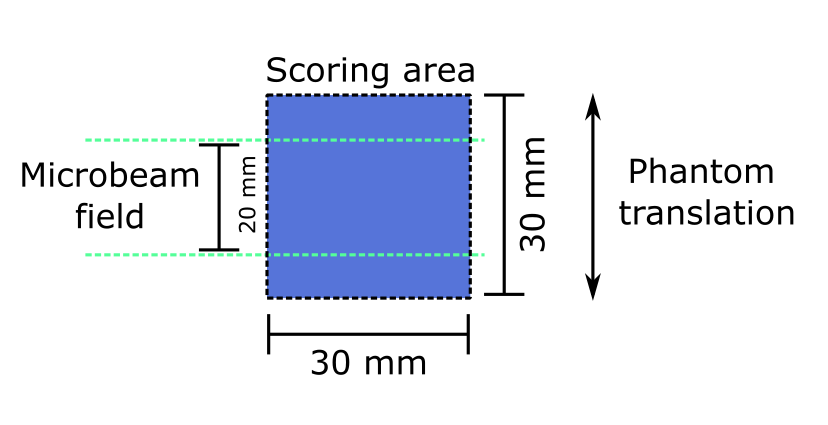}
			\caption{}
			\label{microbeam_field_simulation}
		\end{subfigure}
		\caption{Energy distribution of the microbeams (a), adapted from \cite{Mentzel2022}, and a schematic of the microbeam array irradiation simulation (b) showing a 30x30x30$\,$mm$^3$ water phantom in blue and the edges of the microbeam field in dashed green.}
		\label{microbeams_energy_and_full_field_sim}
	\end{figure}

	The simulation setup for the microbeam array irradiation scenario is shown in Figure (\ref{microbeam_field_simulation}). The microbeams are incident on a small 30x30x30$\,$mm$^3$ water phantom which can be translated to irradiate different sections of the phantom. This size of phantom is chosen because of the current pre-clinical focus on small-scale animal phantoms of MRT research. The microbeam array field obtained from the PSF has an original width of 30$\,$mm and is cropped using a tungsten mask not visible in the schematic to width $20\,$mm as it is a more commonly used configuration. To include dose depositions further away from the beam, the dose is scored in a volume of also 30x30x30$\,$mm$^3$ using voxels of edge length 5$\,\mu$m leading to a total 6000x6000x6000 voxels per simulation for a total of 216 billion voxels. An alternative method for dose scoring is introduced in Section (\ref{section:MacroQuantityScoring}).
	
\begin{figure}[t]
		\centering
		\includegraphics[width=.8\linewidth]{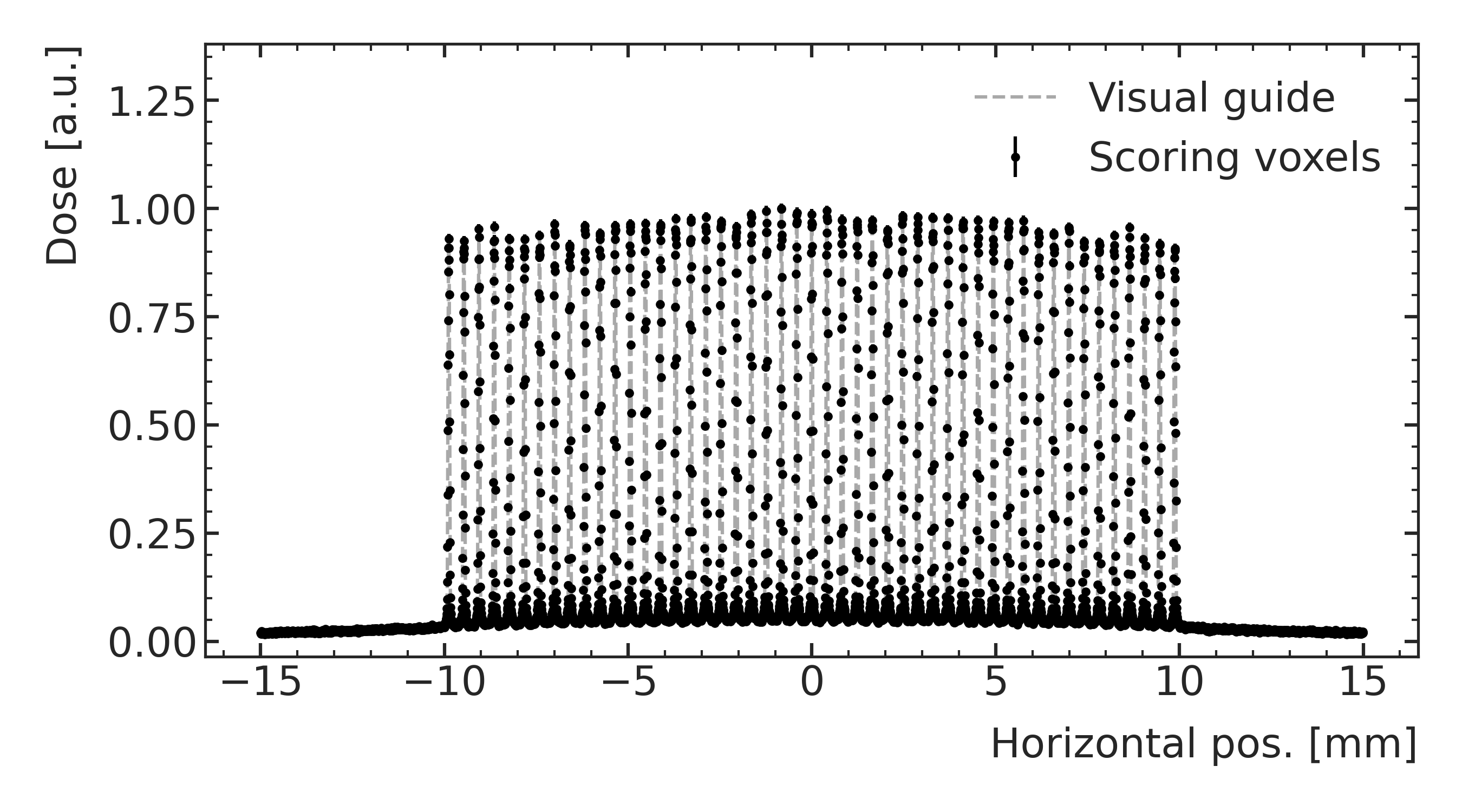}
		\caption{Dose distribution at the centre of the 30x30x30$\,$mm$^3$ water phantom scored with voxels of 5$\,\mu$m edge length (black). A visual guide is given in grey to identify the individual microbeams.}
		\label{dose_distribution_wide}
	\end{figure}

	Figure (\ref{dose_distribution_wide}) shows the dose profile at the centre of the water phantom resulting from the simulation. A grey visual guide is included in the figure to allow for the identification of individual microbeams. The dose deposition at the centre of a microbeam is referred to as \textit{peak dose} while the dose deposition in the middle of two microbeams is referred to as \textit{valley dose}. Differences in peak height are not due to low statistics in the simulations run for this study but originate from the creation of the PSF.
	
	\subsubsection{Single Microbeam Simulation}
	
\begin{figure}[t]
		\centering
		\includegraphics[width=.7\linewidth]{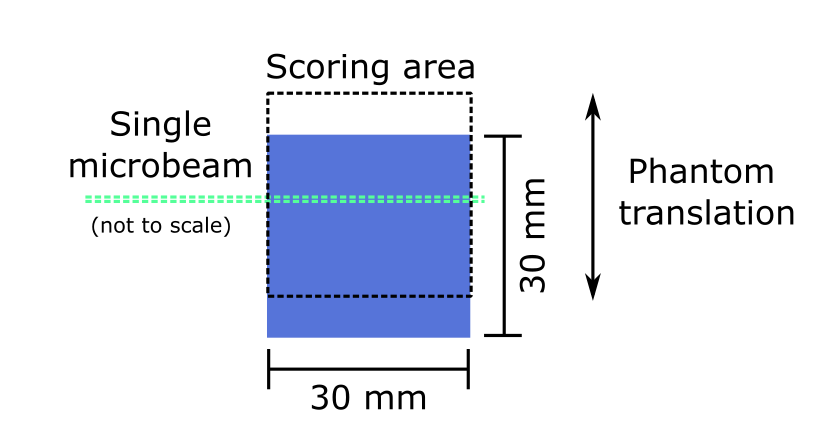}
		\caption{Single microbeam simulation schematic showing a 3x3x3$\,$mm$^3$ water phantom in blue (translated in positive direction) and indicates the position of the microbeam in green.}
		\label{single_microbeam_sim_setup}
	\end{figure}

	To simulate a single individual microbeam, the central microbeam is extracted from the PSF. Figure (\ref{single_microbeam_sim_setup}) shows the schematic for this simulation using the same small-scale water phantom. In addition to the features in Figure (\ref{microbeams_energy_and_full_field_sim}), an exemplary translation of the phantom in positive direction is shown. The scoring area is kept centred around the microbeam containing water and air in that case.

	\subsection{Macro Quantity Scoring\label{section:MacroQuantityScoring}}
	
\begin{figure}[t]
		\centering
		\includegraphics[width=.9\linewidth]{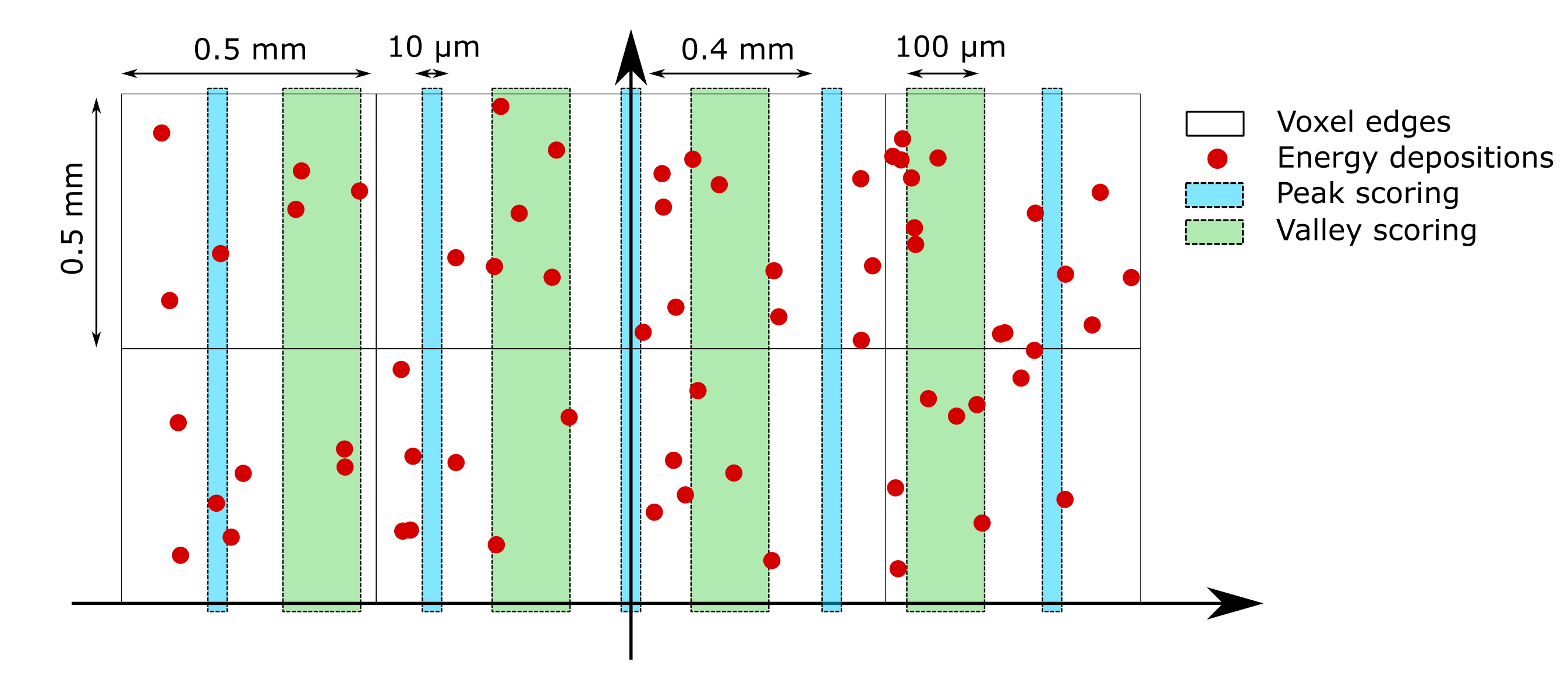}
		\caption{Calculation of peak and valley doses within larger voxels as macroscopic quantities. The voxels are indicated as black squares, the turquoise rectangles indicate that dose depositions (red) which occur within them are counted as peak dose,  the green rectangles indicate counting towards the valley dose.}
		\label{subvoxel_schematic}
	\end{figure}

	In the treatment planning for MRT, a highly resolved dose distribution matrix is often not required. Instead, it is more important to know the distribution of peak and valley doses throughout the phantom. Especially the valley doses are of high importance as those determine the impact on healthy tissue. In contrast to the microscopic dose deposition, those quantities can be computed in a macroscopic manner in larger voxels. Figure (\ref{subvoxel_schematic}) shows a schematic of this approach to MRT data taking for macro voxels of 0.5$\,$mm edge length. During a simulation, the whole scoring volume is separated into \textit{peak scoring volumes}, \textit{valley scoring volumes} and \textit{inactive volumes} which span the whole height of a voxel but only $10\,\mu$m width for the peak scoring and $100\,\mu$m width for the valley scoring (due to shallow dose profile). Dose depositions are counted only towards the \textit{peak dose} or \textit{valley dose} for a given voxel, if they occur within the respective sub-volume.
	
\begin{figure}[t]
		\centering
		\begin{subfigure}[t]{0.41\textwidth}
			\includegraphics[width=\linewidth]{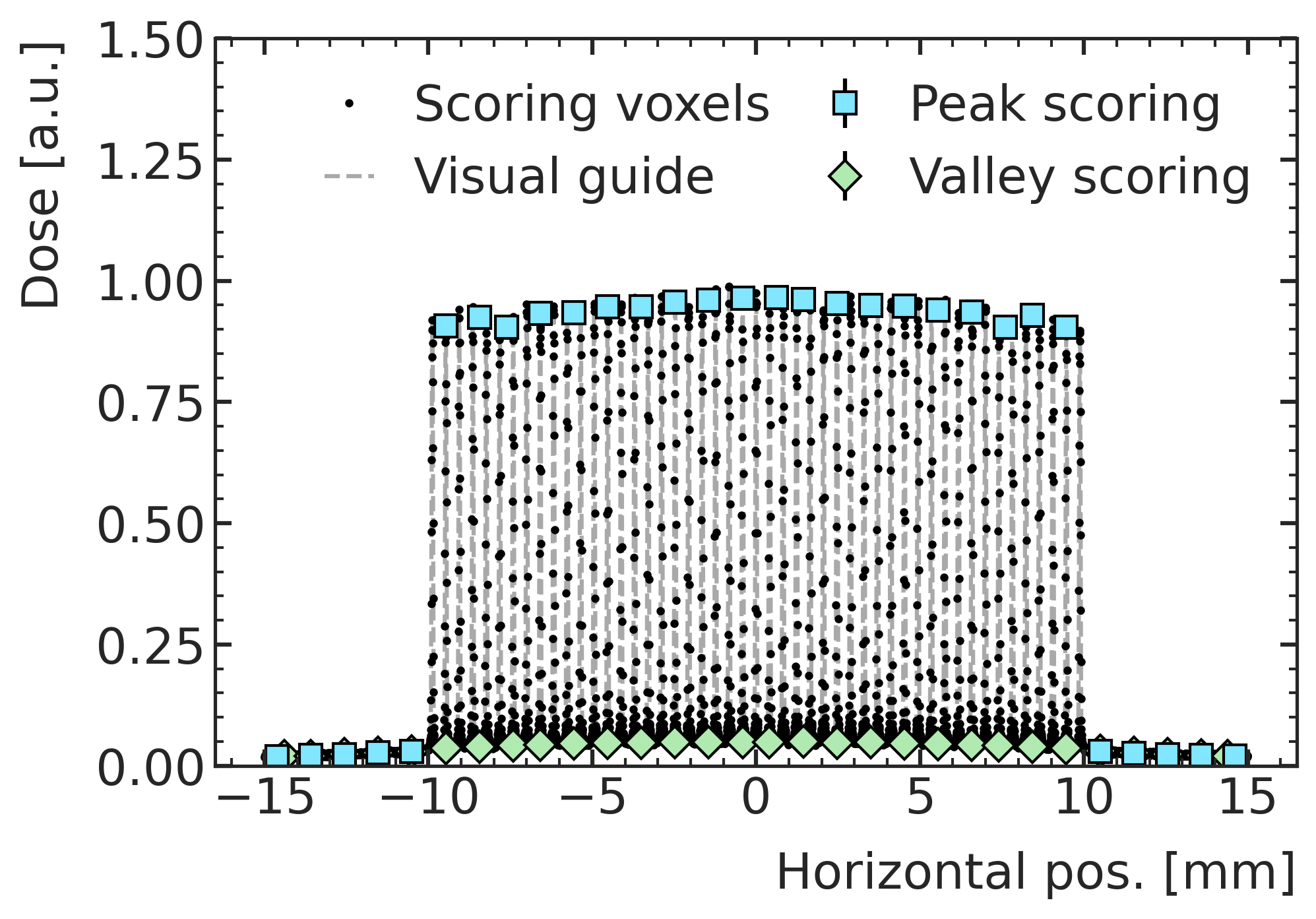}
			\caption{}
			\label{from_micro_to_macro_quantities}
		\end{subfigure}
		\begin{subfigure}[t]{0.48\textwidth}
			\includegraphics[width=\linewidth]{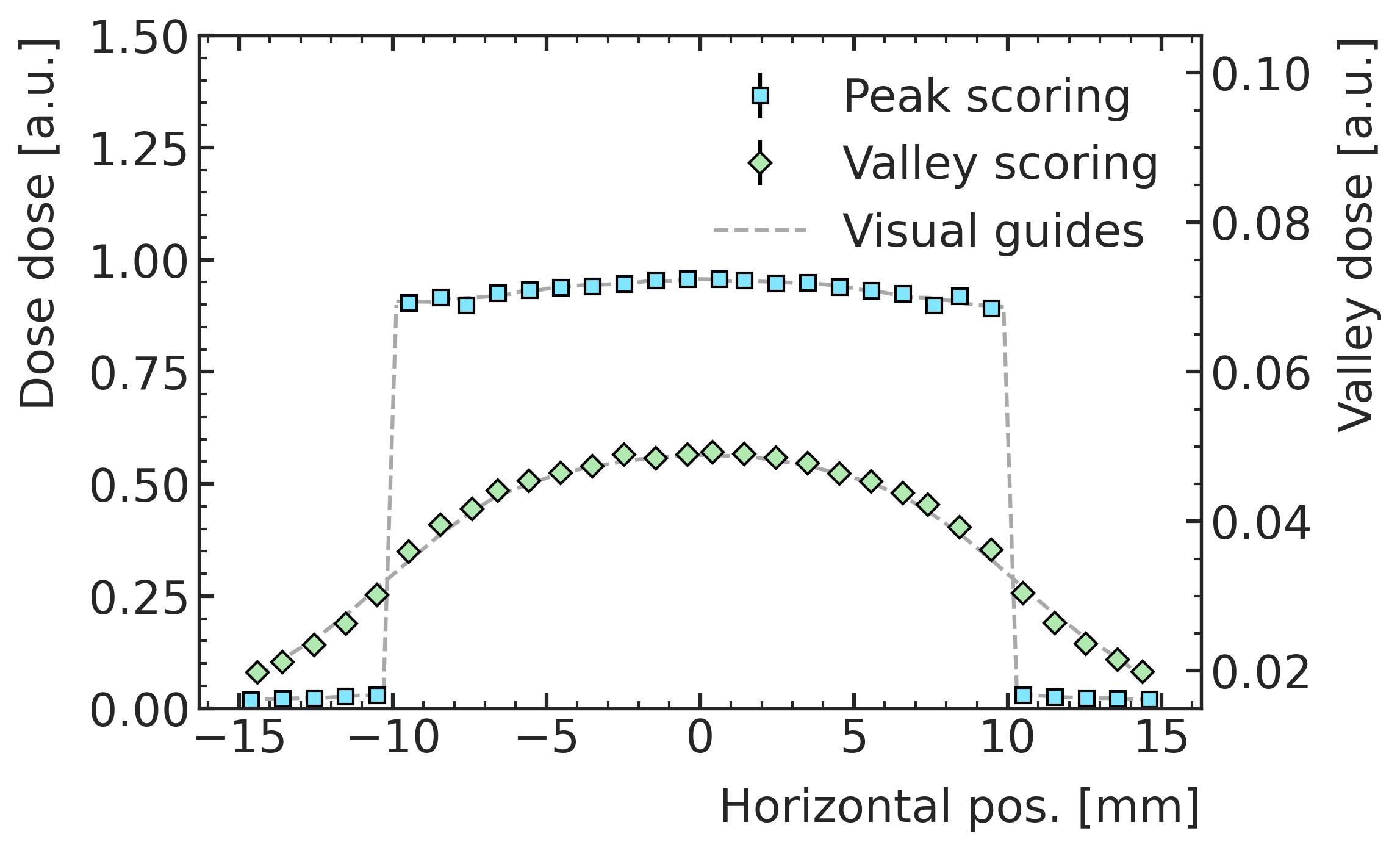}
			\caption{}
			\label{macro_quantities_zoom}
		\end{subfigure}
		\caption{(a) shows the lateral dose profile from Figure (\ref{dose_distribution_wide}) together with the respective peak (turquoise) and valley (green) doses. (a) allows a closer inspection of those macroscopic quantities by showing the same values isolated on individual axis. Visual guides are included in grey.}
		\label{macro_quantities_figure}
	\end{figure}

	This approach simplifies the description of data significantly. Figure (\ref{from_micro_to_macro_quantities}) shows the same lateral dose profile as Figure (\ref{dose_distribution_wide}) but includes the peak and valley doses computed with the presented approach. A more detailed depiction of those in Figure (\ref{macro_quantities_zoom}) shows the lateral peak and valley dose profiles. 
	
\begin{figure}[t]
		\centering
		\begin{subfigure}[t]{0.32\textwidth}
			\includegraphics[width=\linewidth]{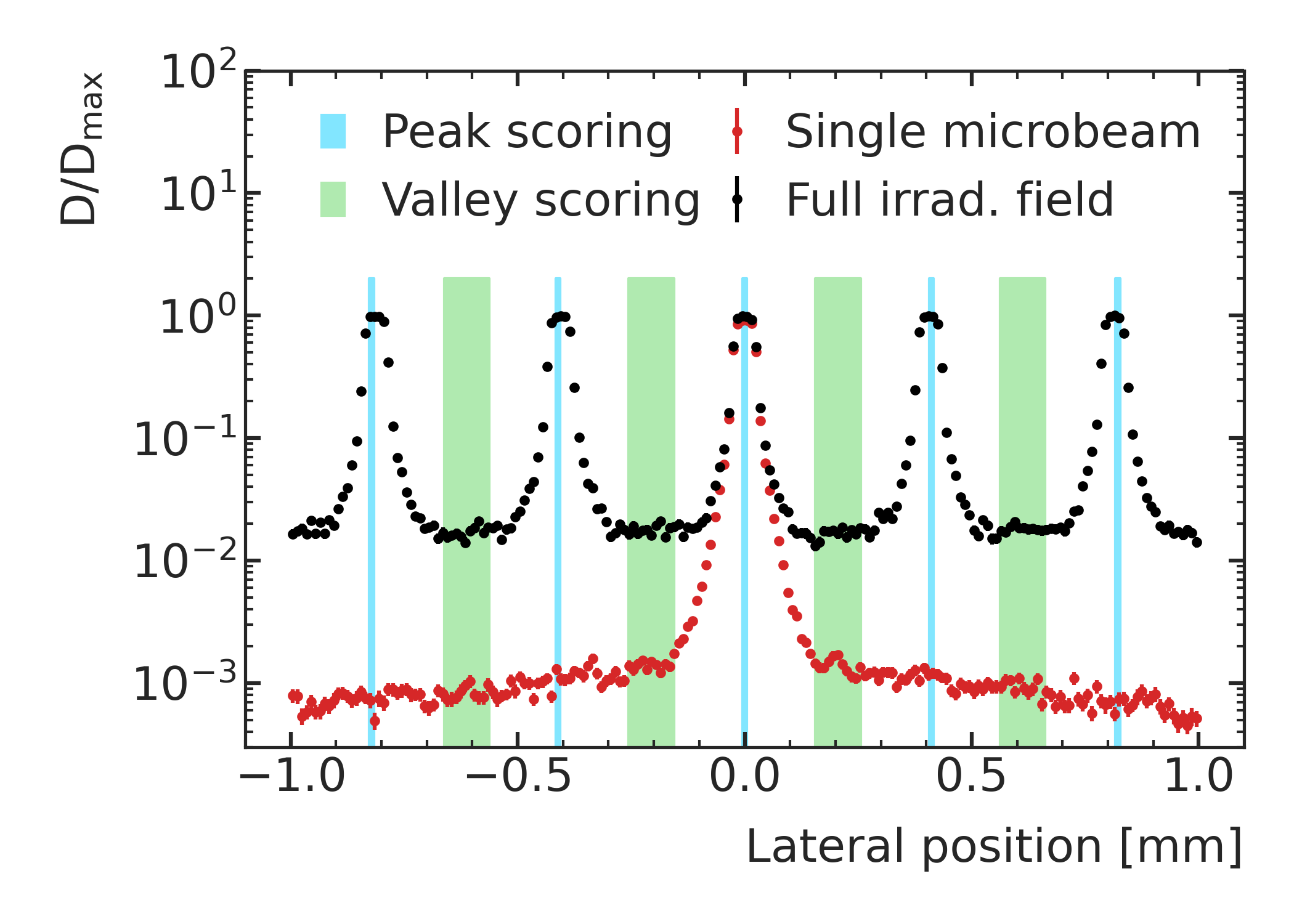}
			\caption{}
			\label{single_beam_peak_valley_scoring}
		\end{subfigure}
		\begin{subfigure}[t]{0.32\textwidth}
			\includegraphics[width=\linewidth]{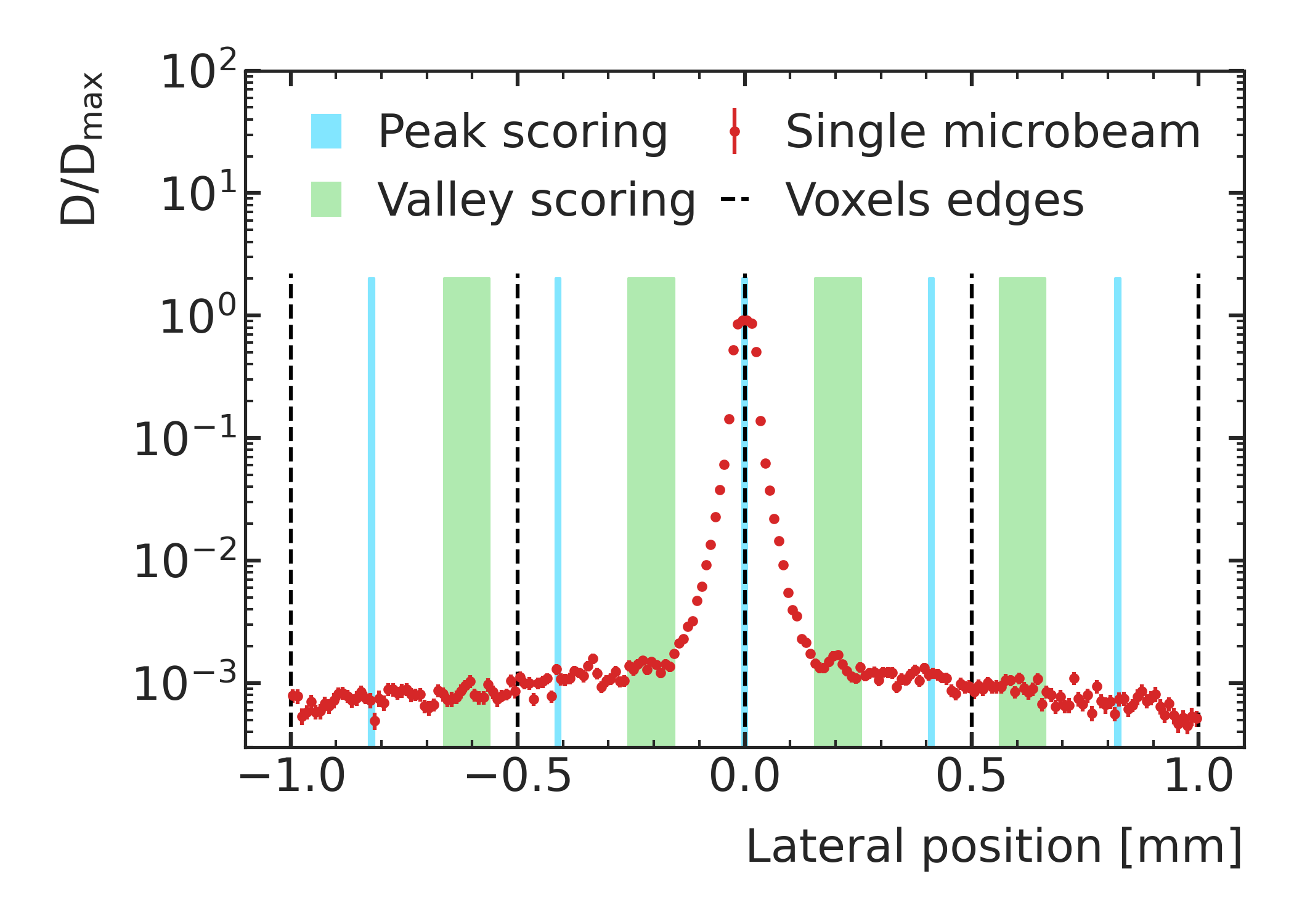}
			\caption{}
			\label{single_beam_peak_valley_scoring_only_mb}
		\end{subfigure}
		\begin{subfigure}[t]{0.32\textwidth}
			\includegraphics[width=\linewidth]{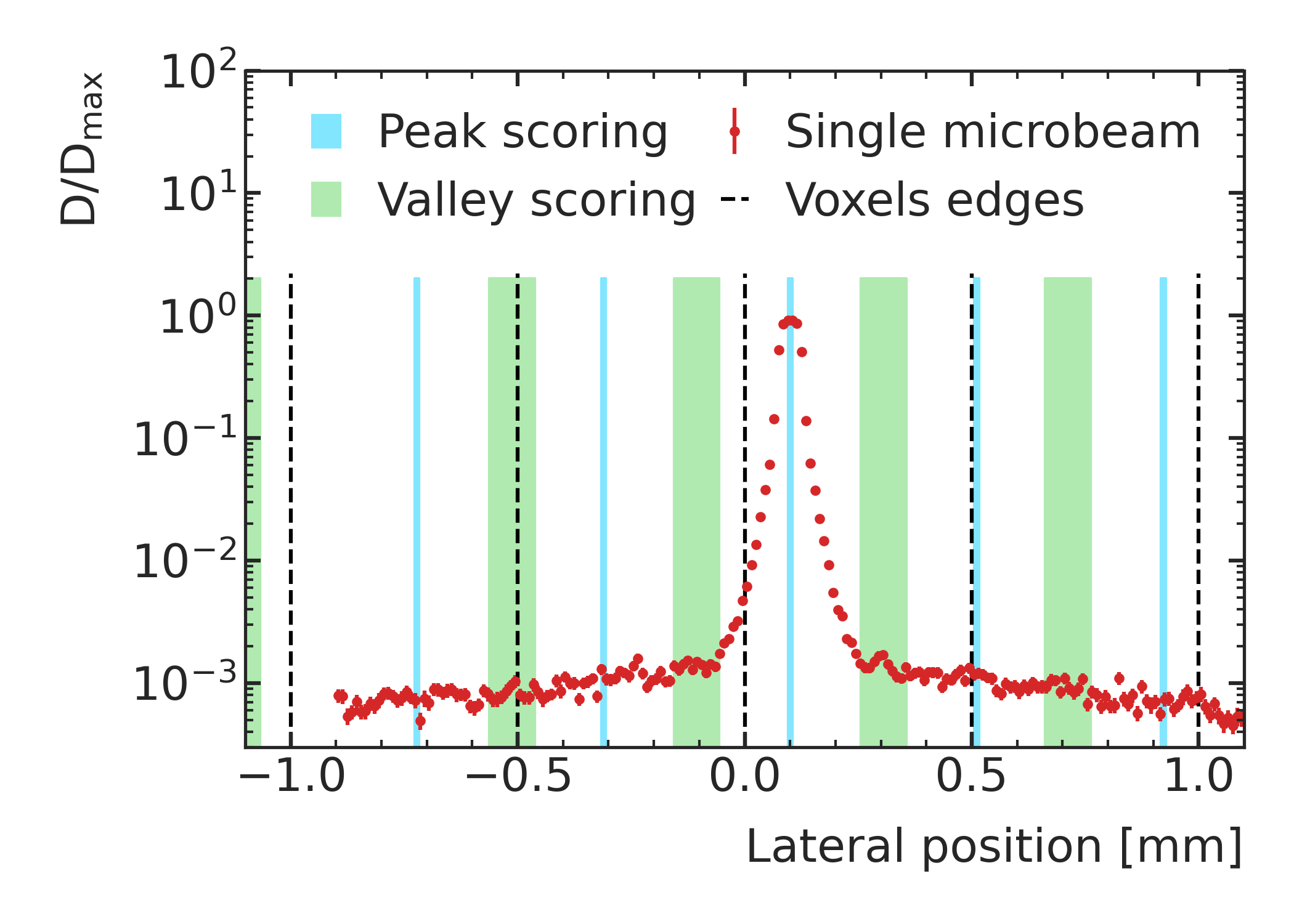}
			\caption{}
			\label{single_beam_peak_valley_scoring_only_mb_shifted}
		\end{subfigure}
		\caption{(a) shows the peak and valley scoring areas in the case of an array of microbeams (black) and a single microbeam (red). (b) and (c) show the translation of a single microbeam and the peak and valley scoring volumes relative to the macro voxels.}
		\label{single_beam_peak_valley_scoring_and_shifted_mb}
	\end{figure}

	The peak and valley doses can also be recorded for a single microbeam by keeping the volumes the same size as in the microbeam array scenario. This is shown in Figure (\ref{single_beam_peak_valley_scoring}). The peak and valley scoring volumes are shifted together with the microbeam while the edges of the macro voxels stay the same. This can be seen in Figure (\ref{single_beam_peak_valley_scoring_only_mb}) and Figure (\ref{single_beam_peak_valley_scoring_only_mb_shifted}).
	
\begin{figure}[t]
		\centering
		\begin{subfigure}[t]{0.35\textwidth}
			\includegraphics[width=\linewidth]{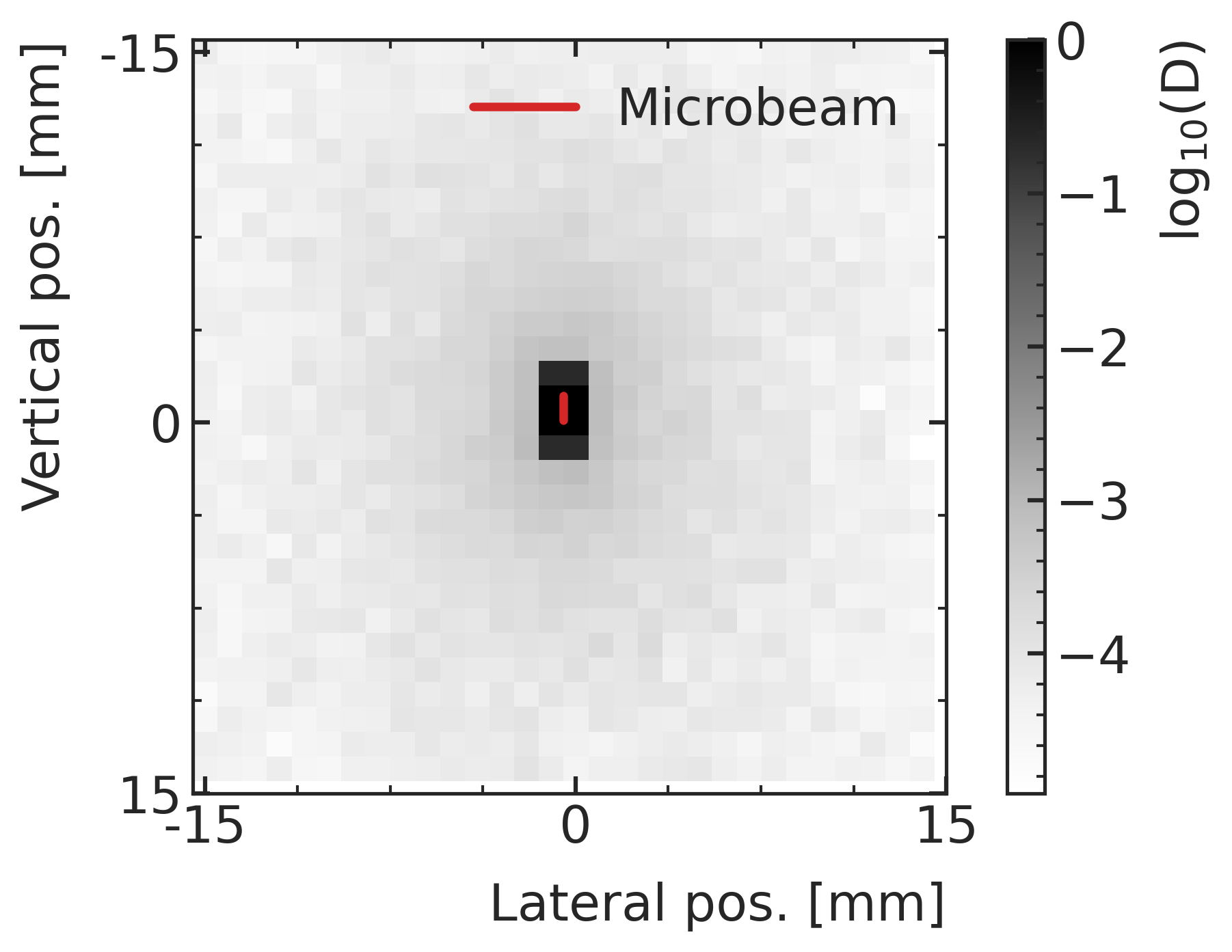}
			\caption{}
			\label{2d_macroVoxelDoses_shifted_beam:a}
		\end{subfigure}
		\begin{subfigure}[t]{0.35\textwidth}
			\includegraphics[width=\linewidth]{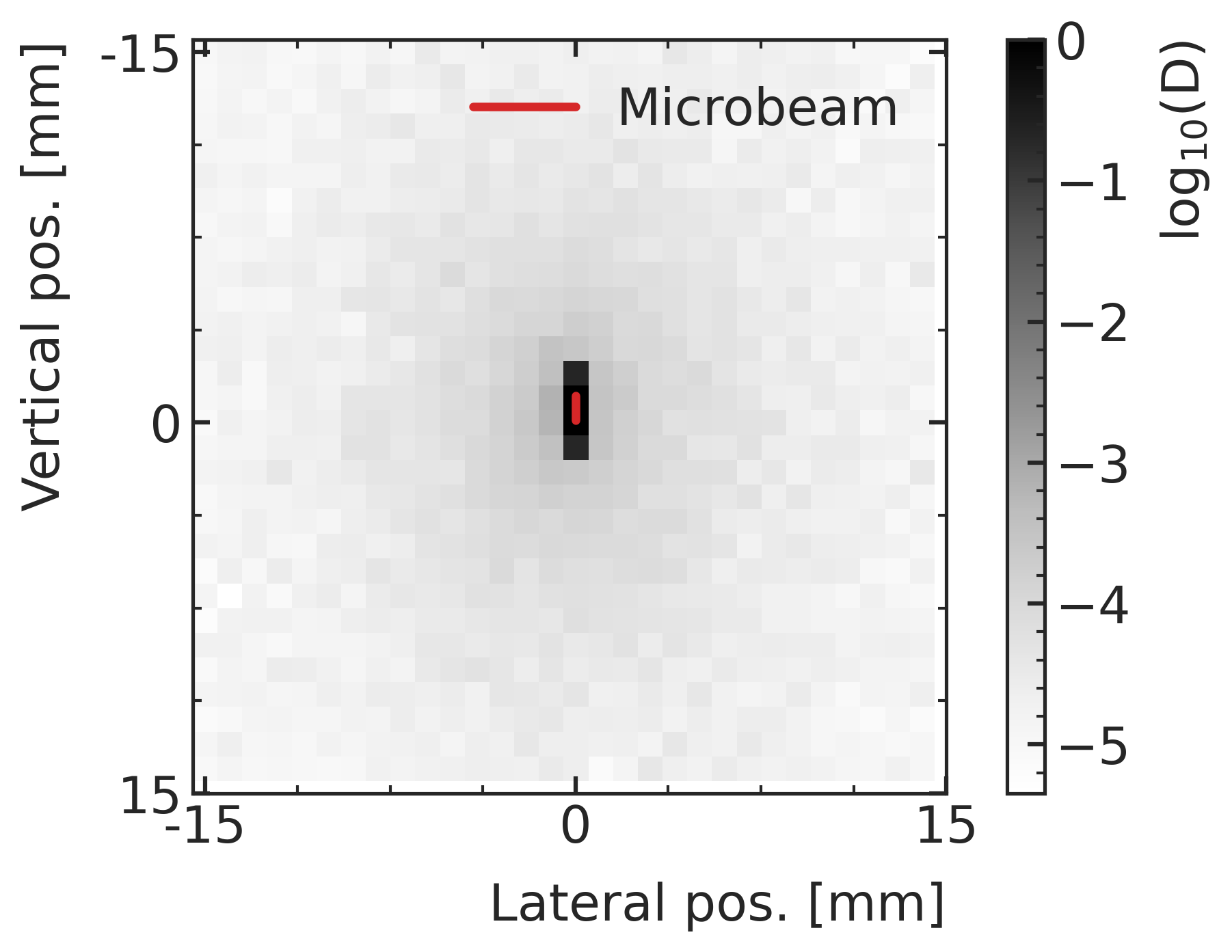}
			\caption{}
			\label{2d_macroVoxelDoses_shifted_beam:b}
		\end{subfigure}
		\caption{Schematic of the peak dose distribution in dependence of the microbeam position relative to the macro voxels.}
		\label{2d_macroVoxelDoses_shifted_beam}
	\end{figure}

	As the dose fall-off is short-ranged compared to the macro voxel size, the dose distributions depend significantly on the initial position of the microbeam relative to the scoring voxels. The effect is shown schematically in Figures (\ref{2d_macroVoxelDoses_shifted_beam:a}) and (\ref{2d_macroVoxelDoses_shifted_beam:b}) where the peak dose per macro voxel is displayed for two different microbeam positions.
	
\begin{figure}[t]
		\centering
		\begin{subfigure}[t]{0.9\textwidth}
			\includegraphics[width=\linewidth]{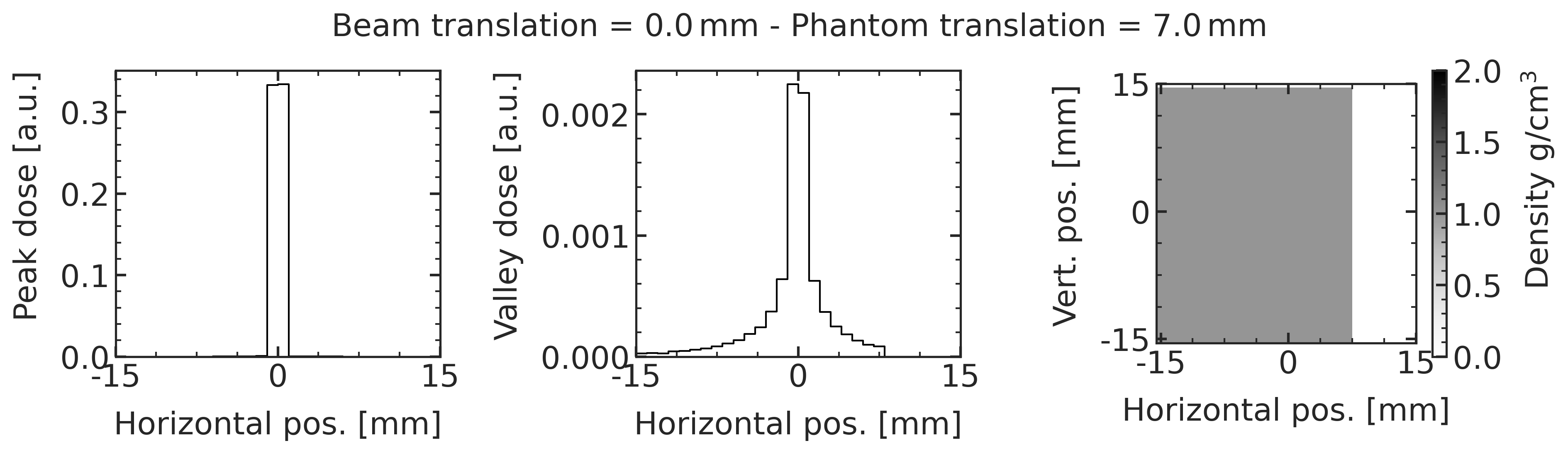}
			\caption{}
			\label{exemplary_data_dose_profiles_and_density:a}
		\end{subfigure}
		\begin{subfigure}[t]{0.9\textwidth}
			\includegraphics[width=\linewidth]{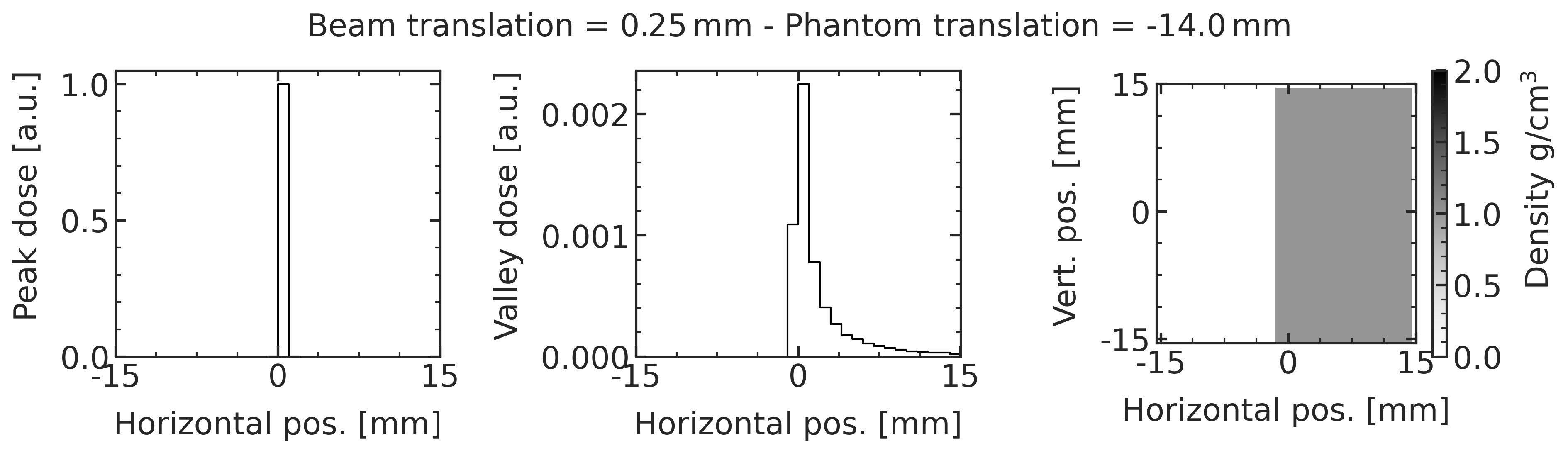}
			\caption{}
			\label{exemplary_data_dose_profiles_and_density:b}
		\end{subfigure}
		\caption{Exemplary lateral peak and valley dose profiles together at the centre of the scoring area together with the density matrix representing the phantom translation for a microbeam translation of 0$\,$mm and a phantom translation of 7$\,$mm (a) and 0.25$\,$mm and -14$\,$mm (b) respectively.}
		\label{exemplary_data_dose_profiles_and_density}
	\end{figure}

	From Figure (\ref{single_beam_peak_valley_scoring}) can also be seen how to superimpose peak and valley doses from individual single microbeams using the macro voxel technique: two macro voxel dose distribution can be added together if the distance between the single microbeams is a multiple of the peak scoring volume distance. In our experiments, the peak doses are always recorded at every $411\,\mu$m. Therefore, a microbeam located e.g. at a global horizontal (lateral) position $y=0\,\mu$m can be added to ones located at $y=\pm411\,\mu$m or $y=\pm822\,\mu$m while a microbeam located at $y=200\,\mu$m could only be added to ones located at e.g. $y=-622\,\mu$m, $y=-211\,\mu$m or $y=+622\,\mu$m. This assumption is acceptable as in a realistic treatment scenario, the spacing between applied microbeams will always be the same for a given multi-slit collimator while the whole array may move relative to the phantom.
	\\
	Figure (\ref{exemplary_data_dose_profiles_and_density}) shows two exemplary lateral peak and valley dose profiles for a single microbeam in the water phantom (including phantom shifts) recorded with macro voxels. In the case of 0$\,$mm beam translation, the centre of the microbeam lies exactly on the border of two macro voxels, splitting the contribution. The peak dose in the central valleys in Figure (\ref{exemplary_data_dose_profiles_and_density:a}) is about 1/3 of the highest peak dose in Figure (\ref{exemplary_data_dose_profiles_and_density:b}) because in the first case, there are two potential peak scoring volumes contributing to each macro voxel which can be seen in Figure (\ref{subvoxel_schematic}). This leads to the occurring energy depositions being divided by more volume. In the second case, the highest voxel contains only one peak scoring volume resulting in no further division of the deposited energy.
	
	\subsection{Machine Learning Model\label{sectino_MLmodel}}
\begin{figure}[t]
		\centering
		\includegraphics[width=\linewidth]{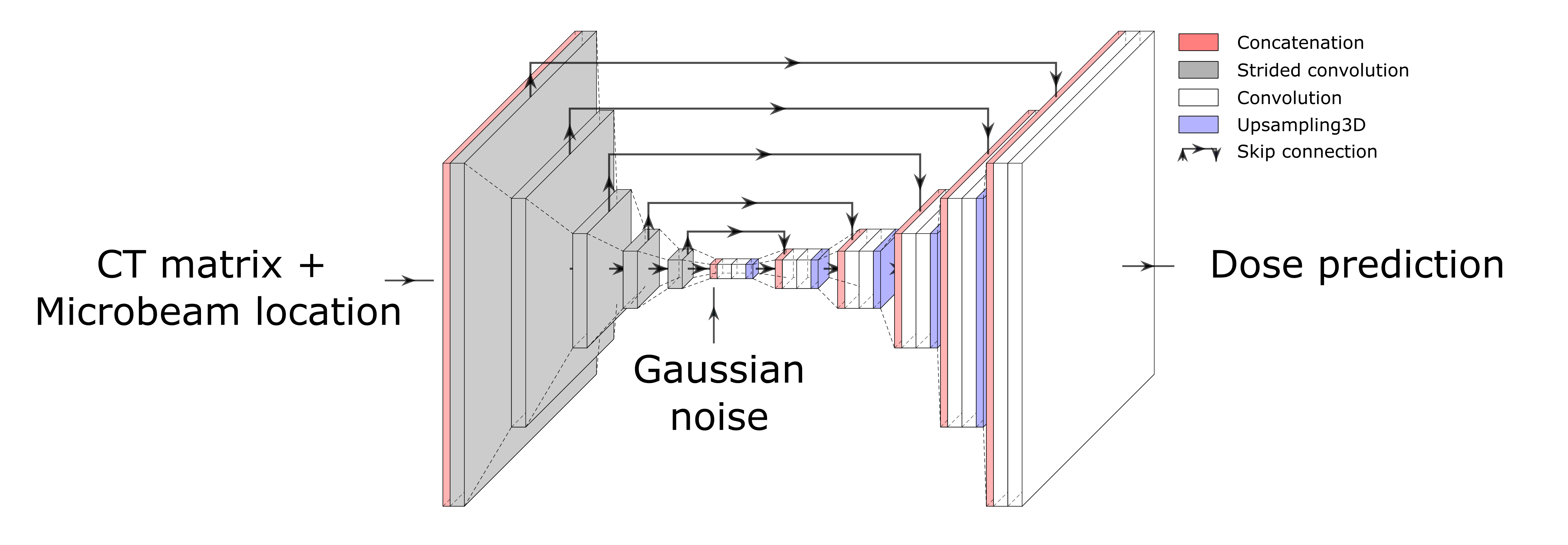}
		\caption{Schematic of the used 3D U-Net model. Adapted from \cite{Mentzel2022}.}
		\label{stripped_generator}
	\end{figure}

	The machine learning model that is trained to predict the macro voxel dose distributions given a microbeam position and a CT/ density matrix is a 3D U-Net adapted from \cite{Mentzel2022} and shown schematically in Figure (\ref{stripped_generator}). The model is implemented using the Keras API \cite{chollet2015keras} for Tensorflow 2.2 \cite{tensorflow2015-whitepaper}. 3D U-Nets comprise a compression stage (left, dark grey), a most compressed so-called bottleneck stage (centre) and a decompression stage (right, red-white-blue). Skip connections in each hierarchical level from compression to decompression allow for feature extraction at different scales and of different complexity. In our model, each convolutional layer comprises 64 filters with a kernel size of 5 for the two first and last levels and kernel size 3 for the other levels. Dropout \cite{Srivastava2014} with a rate of 0.15 is applied after each stage except the output layer.
	
	In contrast to the generative adversarial network model discussed in \cite{Mentzel2022}, the 3D U-Net in this study is trained as regression using the mean squared error as loss function. Two separate networks are trained for the prediction of the peak and the valley macro dose distributions. Both networks are trained with the Adam optimizer \cite{Kingma2015} using a learning rate of $1\cdot 10^{-4}$ and a batch size of 16. 
	
	\subsection{Machine Learning Training Data}
\begin{figure}[t]
		\centering
		\includegraphics[width=.6\linewidth]{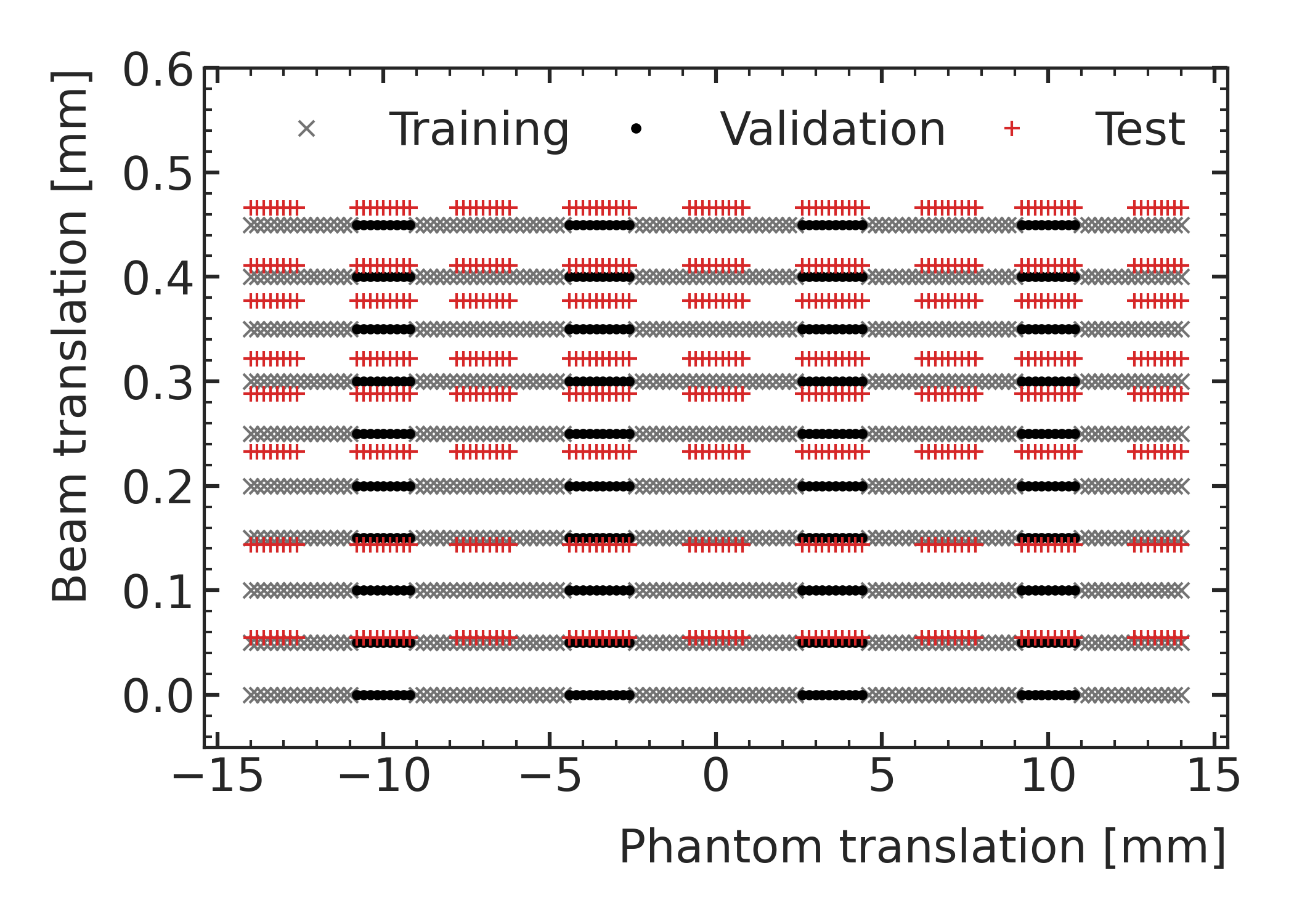}
		\caption{Training (grey), validation (black) and test (red) data points as a function of the phantom and beam translation.}
		\label{train_val_test_data}
	\end{figure}

	Figure (\ref{train_val_test_data}) shows the split of data points into training, validation and test samples used for the study. The data points differ in phantom translation $\Delta_\mathrm{P}$ and beam translation $\Delta_\mathrm{B}$. Data samples from phantom translations of $-11\,\mathrm{mm} <= \Delta_\mathrm{P} <= -9\,\mathrm{mm}$, $-4.5\,\mathrm{mm} <= \Delta_\mathrm{P} <= -2.5\,\mathrm{mm}$, $2.5\,\mathrm{mm} <= \Delta_\mathrm{P} <= 4.5\,\mathrm{mm}$ and $9\,\mathrm{mm} < \Delta_\mathrm{P} < 11\,\mathrm{mm}$ are used as validation samples meaning that they are not presented to the network for training but only for performance evaluation during the training to compare different ML model configurations. 
	\\
	In addition, a test set of data samples is held out during the entire search for an optimal model which is only used for performance evaluation at the end of the study to allow for a minimal bias. The test set comprises of several phantom translations at beam translations of multiples of $411\,\mu$m (modulo $500\,\mu$m) instead of the regular grid used for training and validation data. The test data, therefore, resembles the beam predictions needed for predicting microbeams that can be superimposed to build a microbeam irradiation array. 
	\\
	For this proof of concept, the simulated data is reduced in size to 30x30x30 voxels of edge length 1$\,$mm instead of 60x60x60 voxels of edge length $0.5\,$mm for the training of the ML model.
	
	\section{Results\label{section:results}}
	\subsection{Superposition Of Single Microbeams}
\begin{figure}[t]
		\centering
		\begin{subfigure}[t]{0.45\textwidth}
			\includegraphics[width=\linewidth]{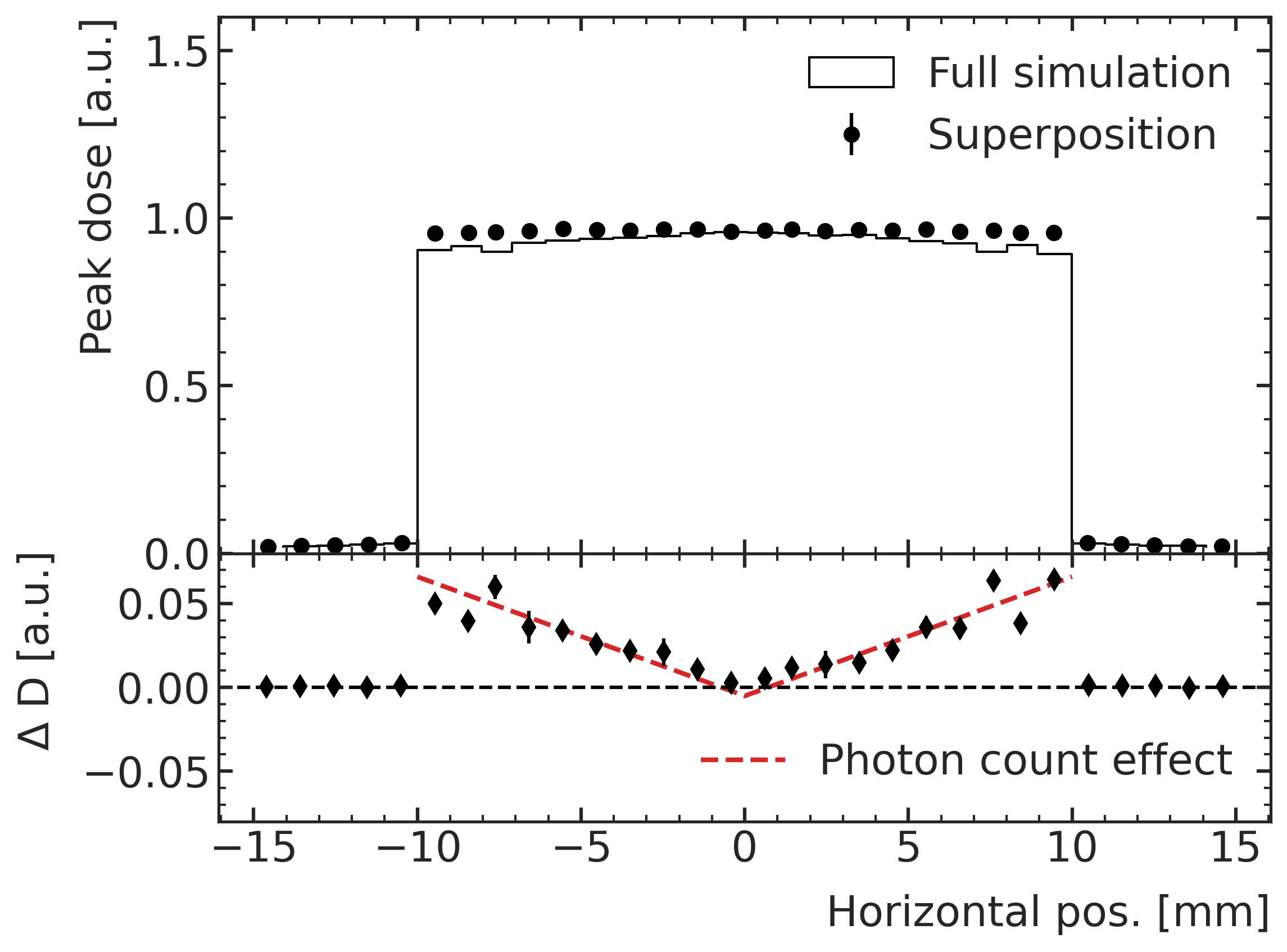}
			\caption{}
			\label{superposition_peak_and_valley_doses:a}
		\end{subfigure}
		\begin{subfigure}[t]{0.45\textwidth}
			\includegraphics[width=\linewidth]{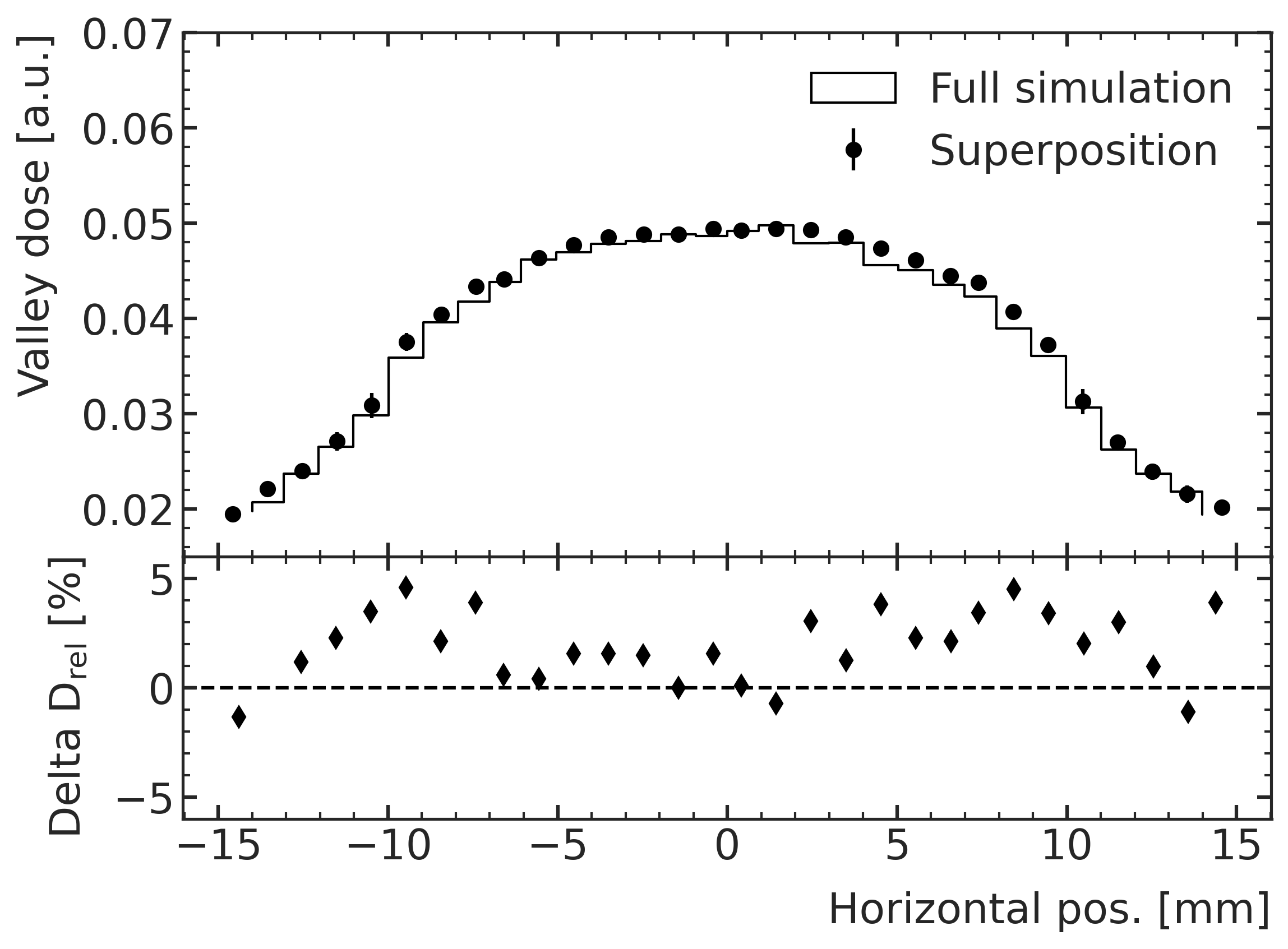}
			\caption{}
			\label{superposition_peak_and_valley_doses:c}
		\end{subfigure}
		\caption{Comparison between peak (a) and valley (b) doses of a microbeam array from superposition of single microbeams and the full field simulation. The red line in (a) indicates the effect of the photon number per microbeam.}
		\label{superposition_peak_and_valley_doses}
	\end{figure}

	Figure (\ref{superposition_peak_and_valley_doses}) shows the peak and valley doses for a 20$\,$mm wide microbeam array field. The red markers indicate the results obtained from the full microbeam array simulation while the data indicated by the black markers is obtained from the superposition of multiple translated simulations of the central microbeam. It can be seen, that the full field simulation shows a lateral roll-off while the peak doses of the superimposed field are more even. Together with the peak doses, a black line indicates the effect of the number of photons that comprise the individual microbeams. Due to beam diversion, the multi-slit collimator crops more photons from the microbeams further outside from the field centre leaving them with less photons contributing to the peak doses. This effect will have to be corrected for when using superimposed single microbeam dose distributions by scaling the peak doses accordingly. In Figure (\ref{superposition_peak_and_valley_doses:c}) it can be seen that the valley doses systematically lie 0-5\% lower for the full simulation over the whole width of the field. Whether the correction of the peak doses applied to the single beam valley doses would suffice to correct this difference will be required to be investigated in future studies.

	\subsection{Peak And Valley Dose Predictions}
\begin{figure}[t]
		\centering
		\begin{subfigure}[t]{0.45\textwidth}
			\includegraphics[width=\linewidth]{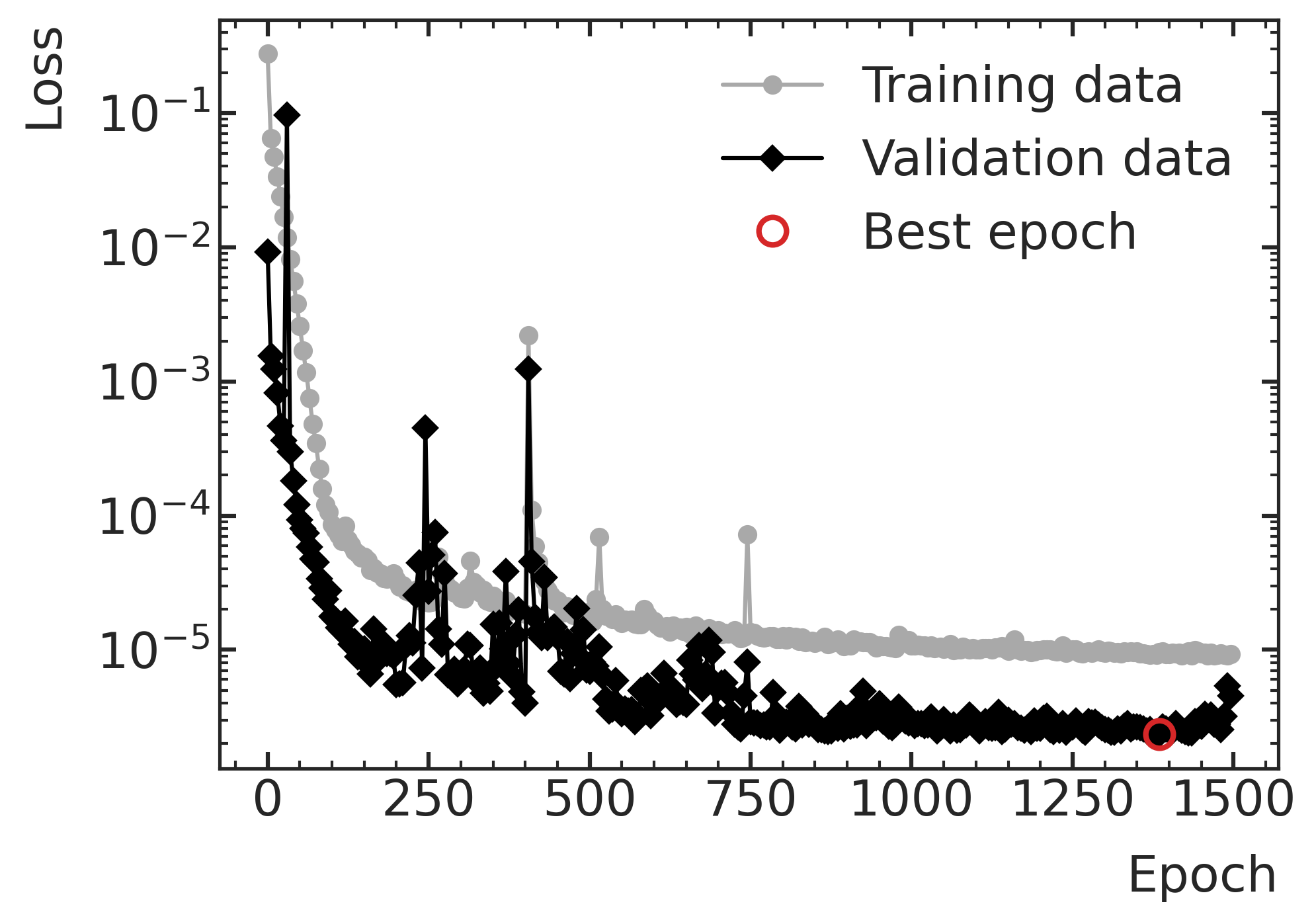}
			\caption{}
			\label{loss_curves:a}
		\end{subfigure}
		\begin{subfigure}[t]{0.45\textwidth}
			\includegraphics[width=\linewidth]{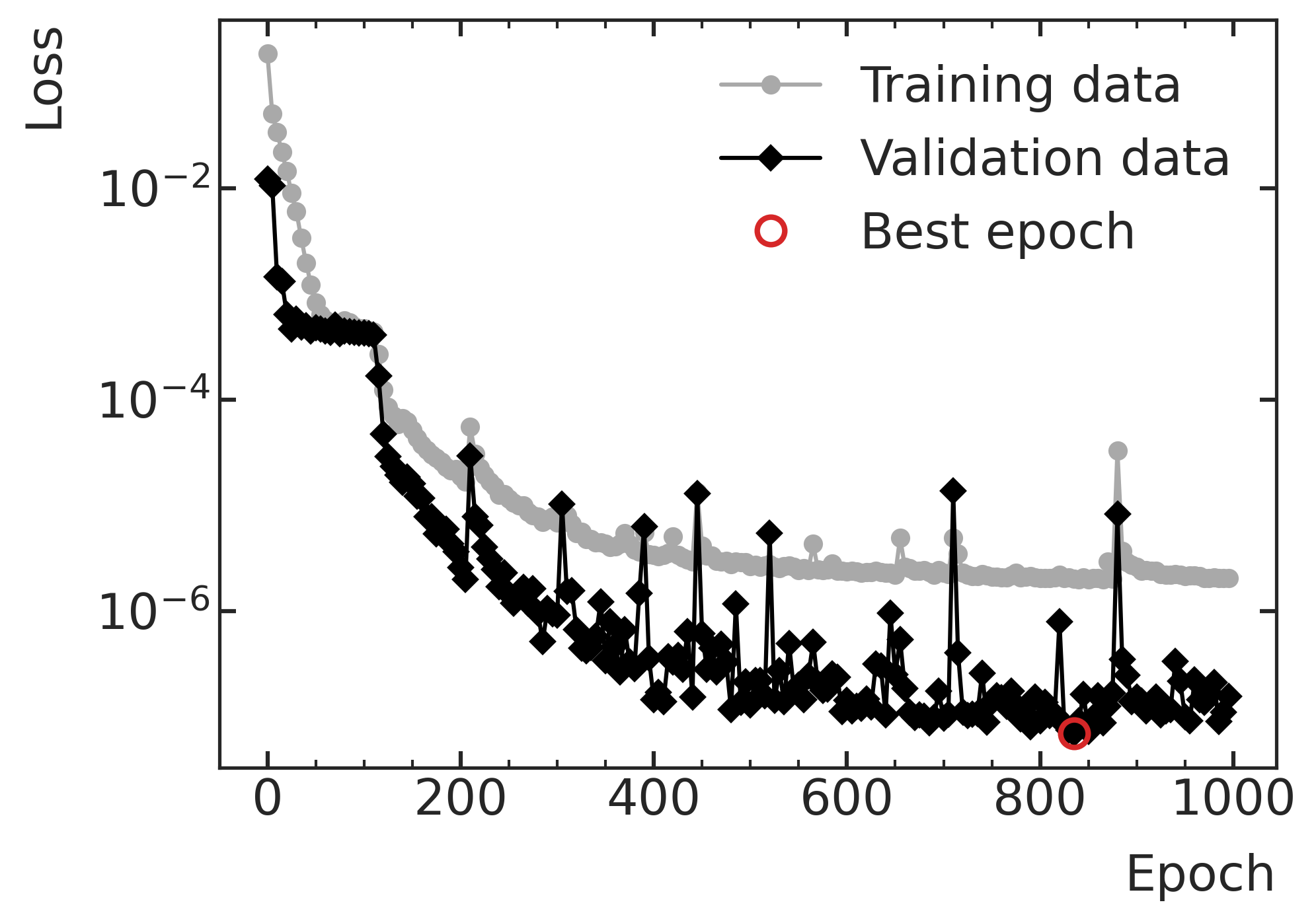}
			\caption{}
			\label{loss_curves:b}
		\end{subfigure}
		\caption{Training (grey) and validation (black) loss (mean squared error) for the valley (a) and peak (b) dose network. The red circle indicates the epoch with the lowest validation error respectively.}
		\label{loss_curves}
	\end{figure}

	Each 3D U-Net is trained until the validation loss is not further reduced. The minimum validation loss for the valley dose prediction is achieved after 1306 epochs, the minimum for the peak dose distribution after 839 epochs. The corresponding loss curves are shown in Figure (\ref{loss_curves}). The validation loss lies systematically below the training loss because especially predictions with large phantom translations with the water-air transition being closer to the microbeam centre induce large errors. The edge cases are not included in the validation set, therefore the loss stays lower.
	
	To investigate the generalization of the training, all training data, validation data and now also the test data is predicted. The prediction time with the used \textit{NVidia GForce 1080Ti} averages $t=116\pm4\,$ms for individual sequential predictions. In batch mode, predicting multiple configurations simultaneously, this can be significantly decreased. Using the training batch size of 16, those samples can be predicted within $t_\mathrm{batch}=241.8\pm0.3\,$ms on average resulting in a prediction time per sample of $t_\mathrm{sample}\approx15\,$ms. Based on this, a full 20x20$\,$mm$^2$ microbeam array field consisting of 70 single microbeams, each 20$\,$mm high would take approximately 40 seconds (20$\,$s for peak and valley dose each) to compute. It can be expected that a more general model for more complex geometries would not result in slower prediction times.
	\begin{figure}[t]
		\centering
		\begin{subfigure}[t]{0.45\textwidth}
			\includegraphics[width=\linewidth]{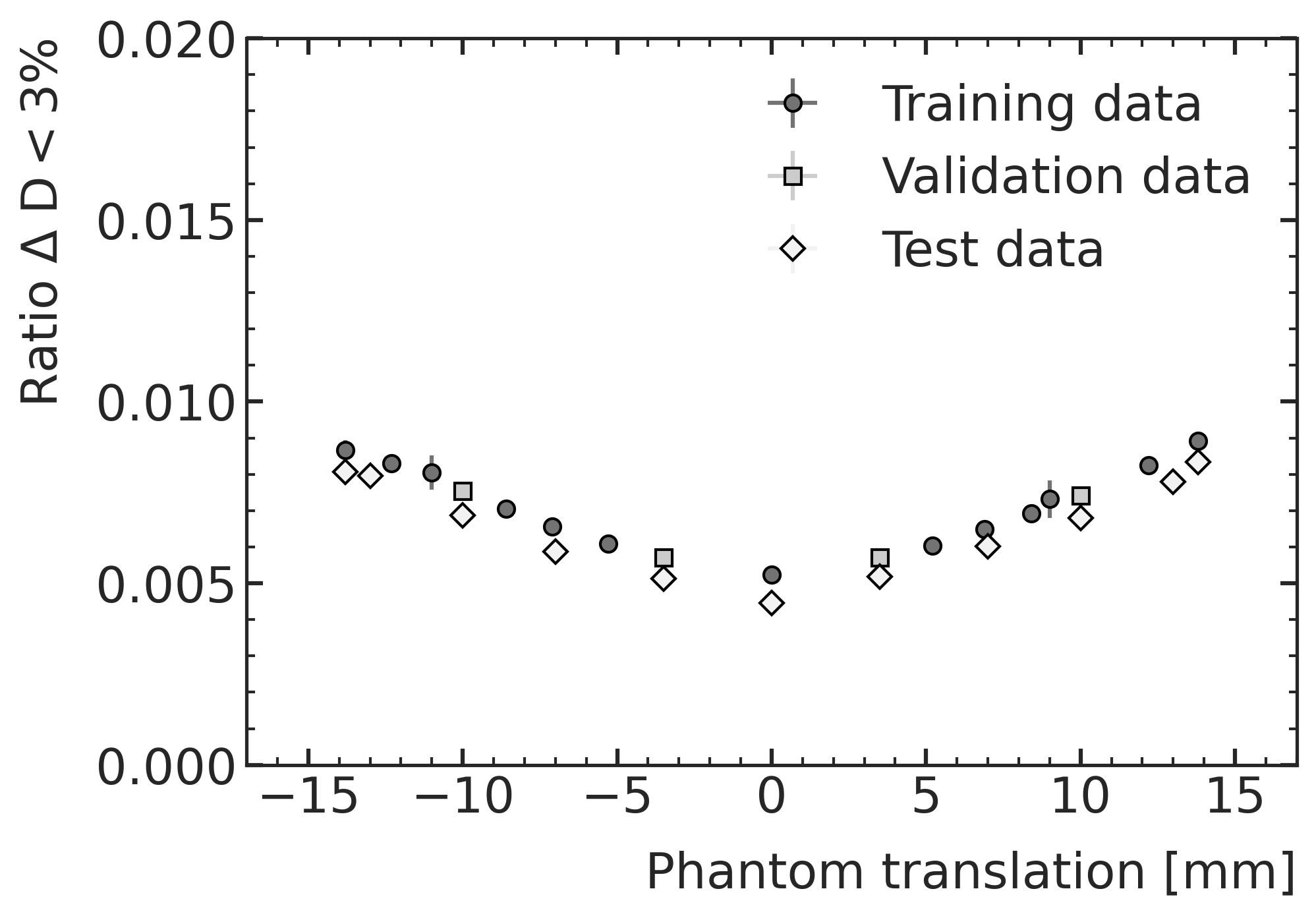}
			\caption{}
			\label{generalization_metrics:a}
		\end{subfigure}
		\begin{subfigure}[t]{0.45\textwidth}
			\includegraphics[width=\linewidth]{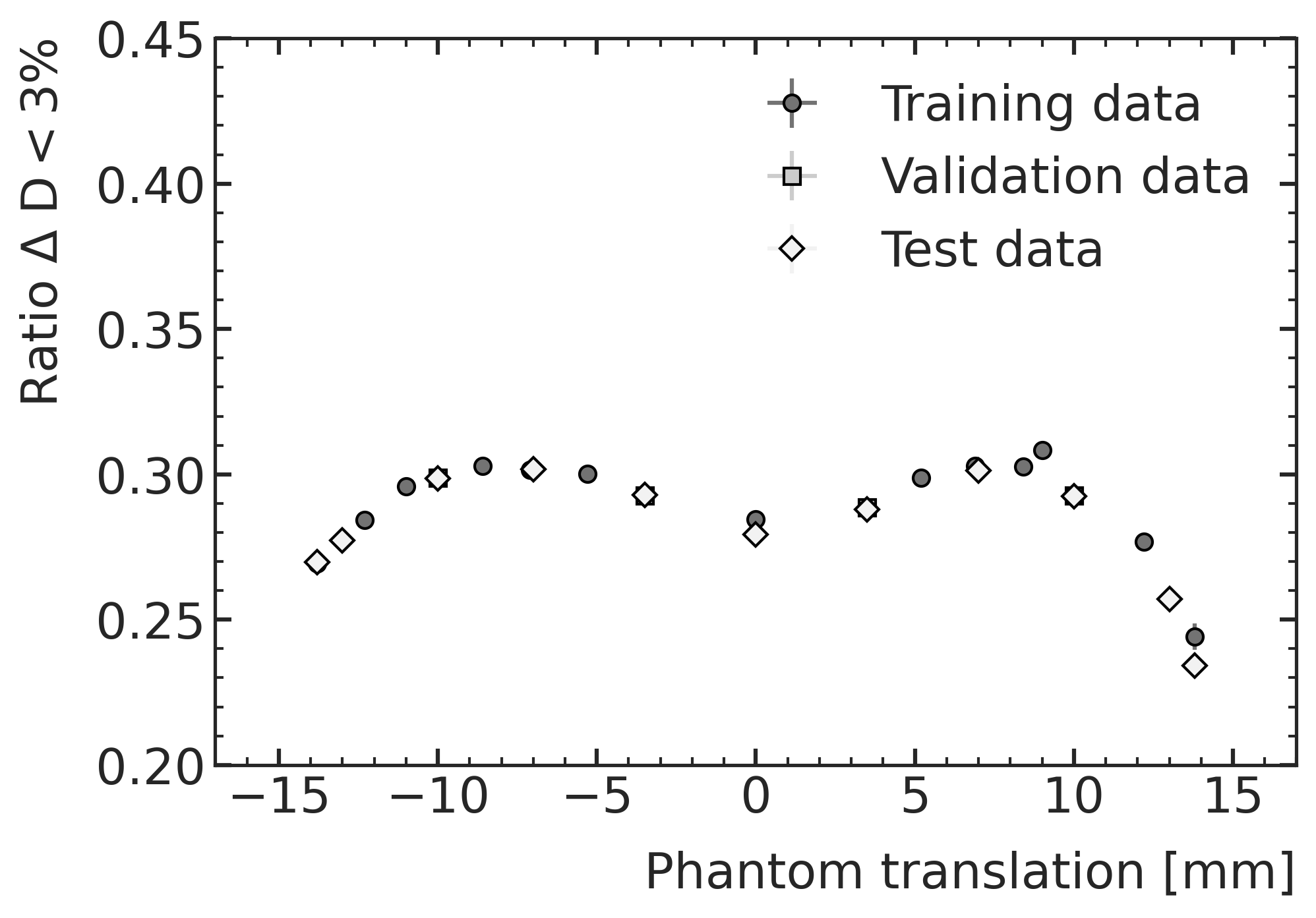}
			\caption{}
			\label{generalization_metrics:b}
		\end{subfigure}
		\caption{Ratio of voxels deviating less than 3\% from the MC ground truth in the peak (a) and valley (b) predictions of training, validation and test data.}
		\label{generalization_metrics}
	\end{figure}
	
\begin{figure}[t]
		\centering
		\begin{subfigure}[t]{0.32\textwidth}
			\includegraphics[width=\linewidth]{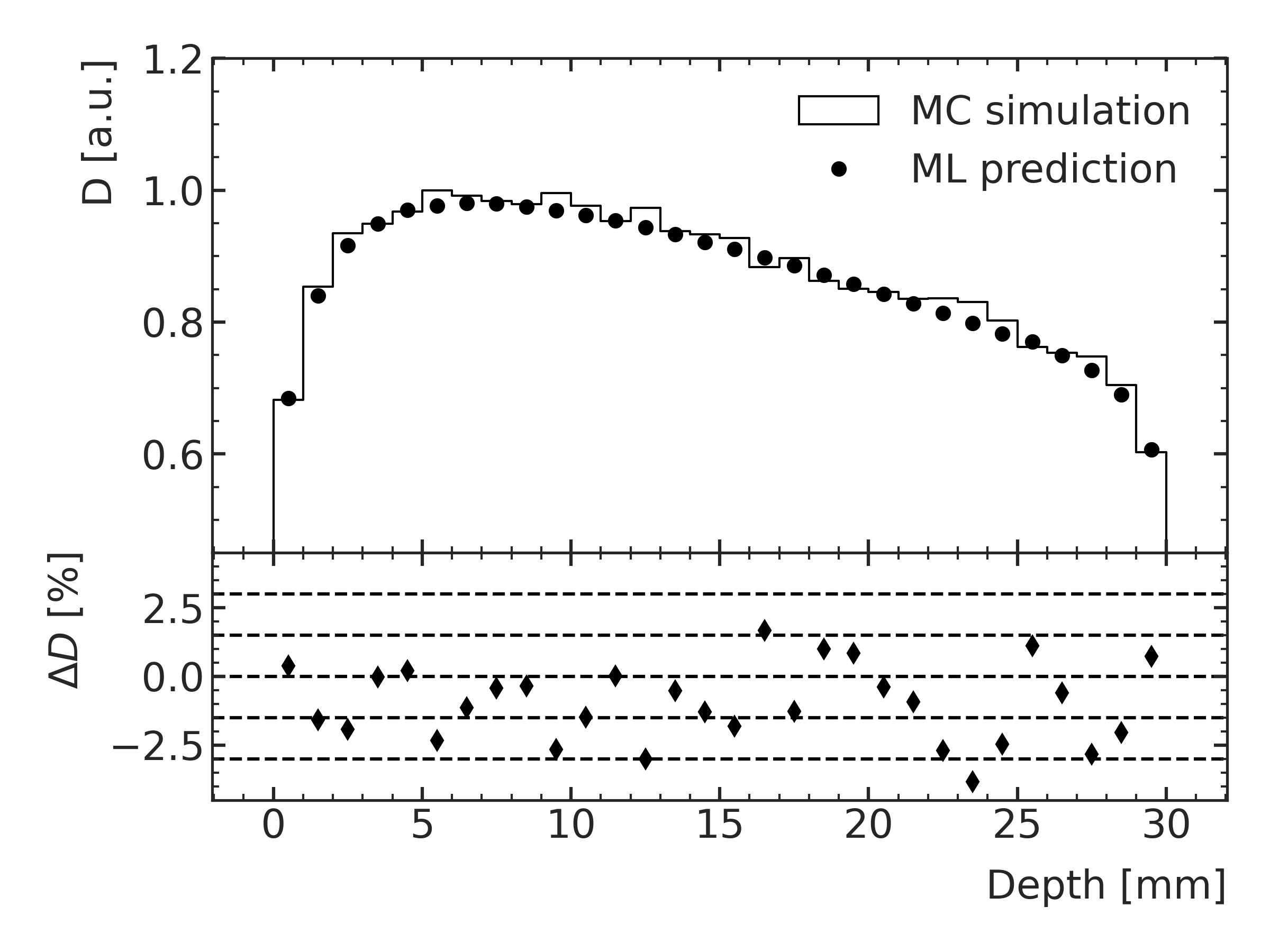}
			\caption{}
			\label{valley_predictions:a}
		\end{subfigure}
		\begin{subfigure}[t]{0.32\textwidth}
			\includegraphics[width=\linewidth]{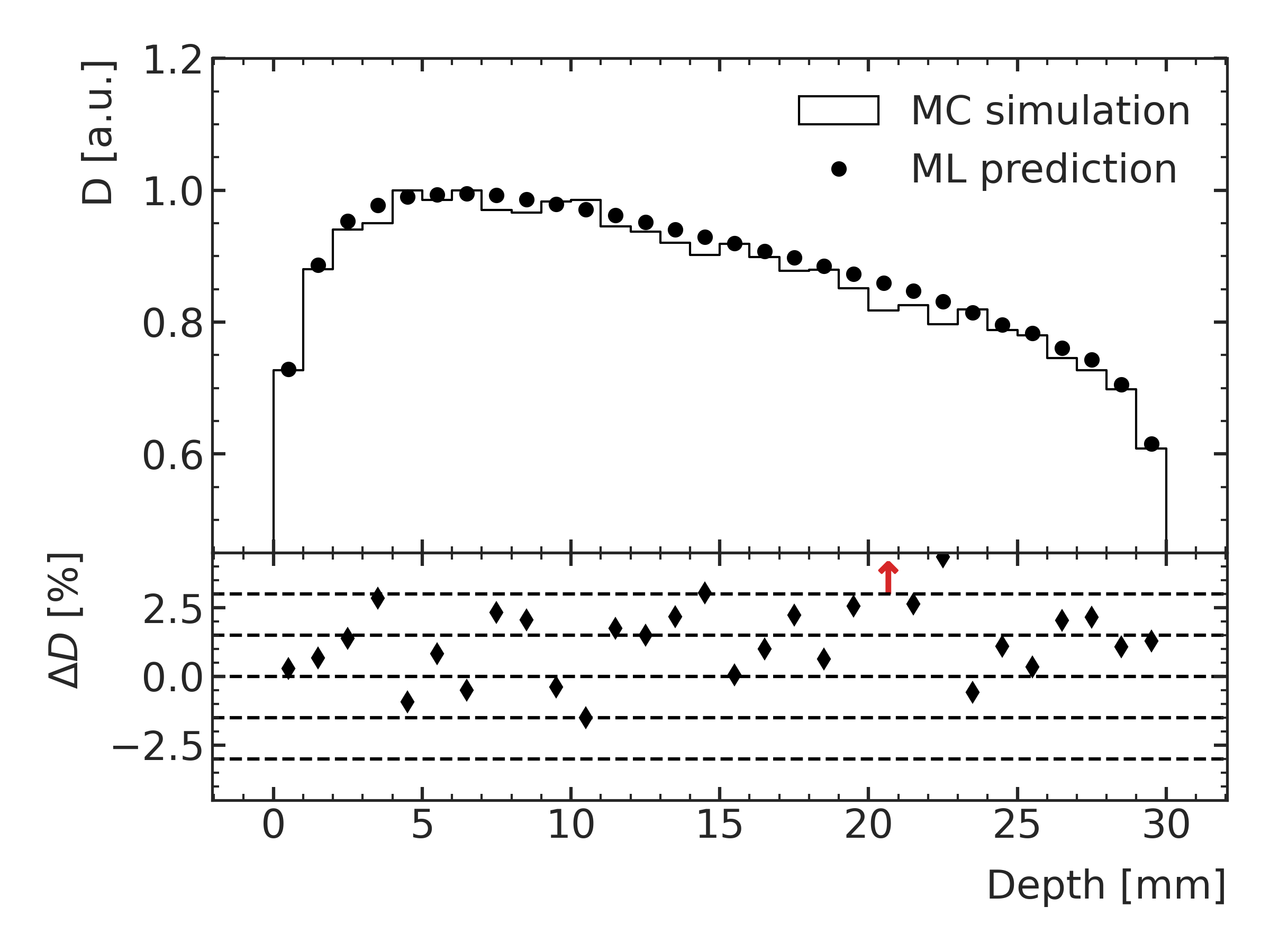}
			\caption{}
			\label{valley_predictions:b}
		\end{subfigure}
		\begin{subfigure}[t]{0.32\textwidth}
			\includegraphics[width=\linewidth]{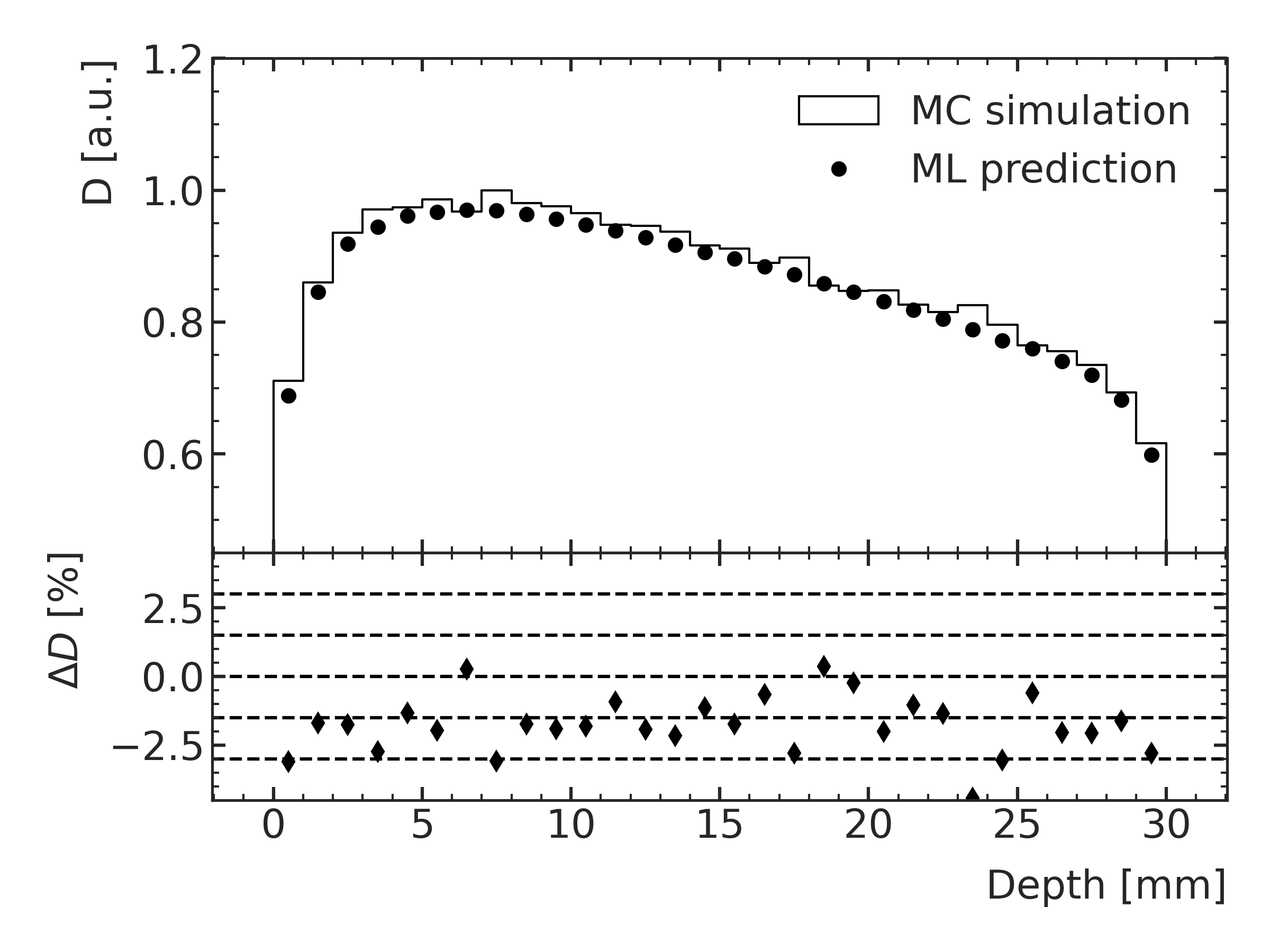}
			\caption{}
			\label{valley_predictions:c}
		\end{subfigure}
		\begin{subfigure}[t]{0.32\textwidth}
			\includegraphics[width=\linewidth]{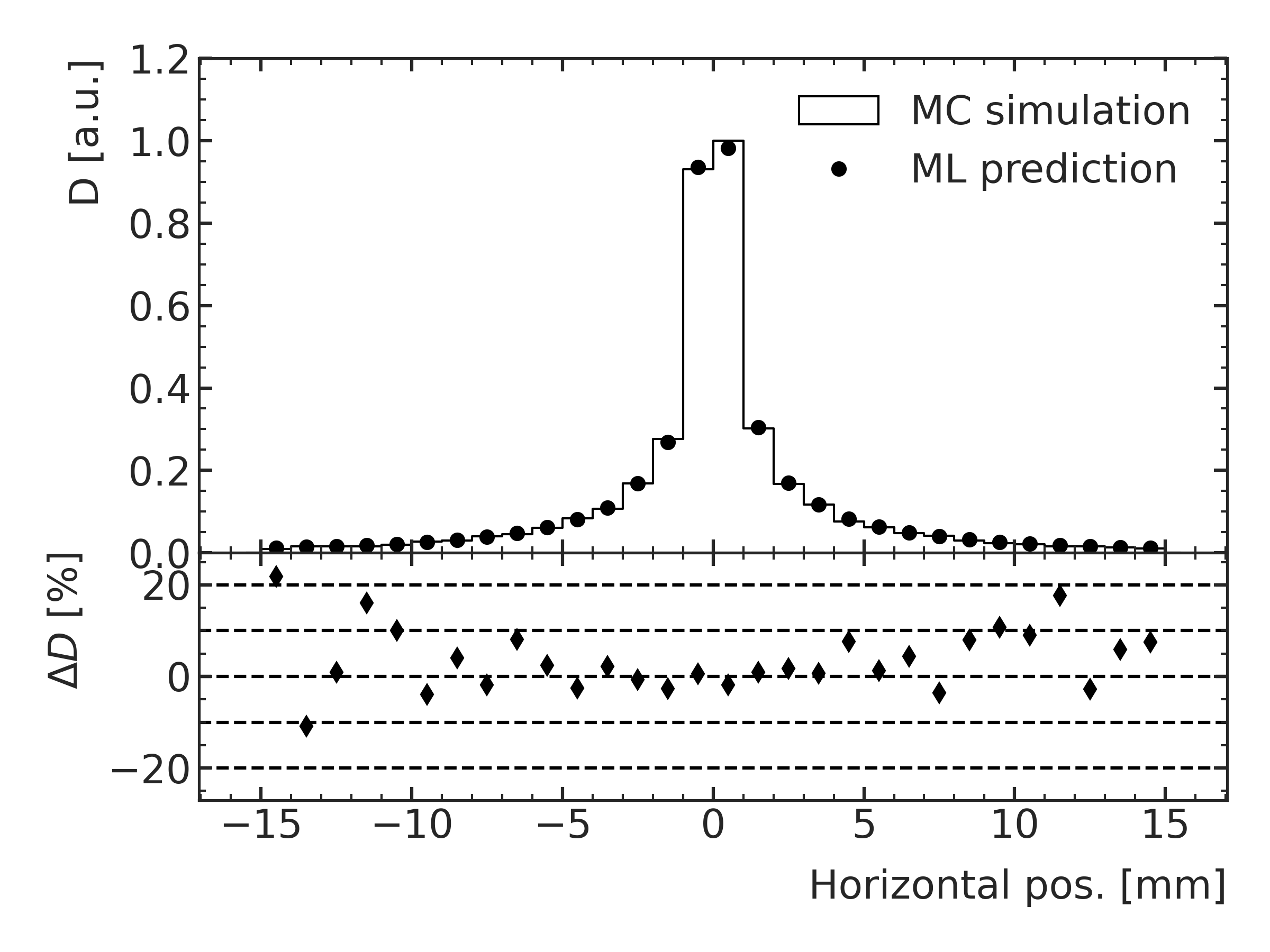}
			\caption{}
			\label{valley_predictions:d}
		\end{subfigure}
		\begin{subfigure}[t]{0.32\textwidth}
			\includegraphics[width=\linewidth]{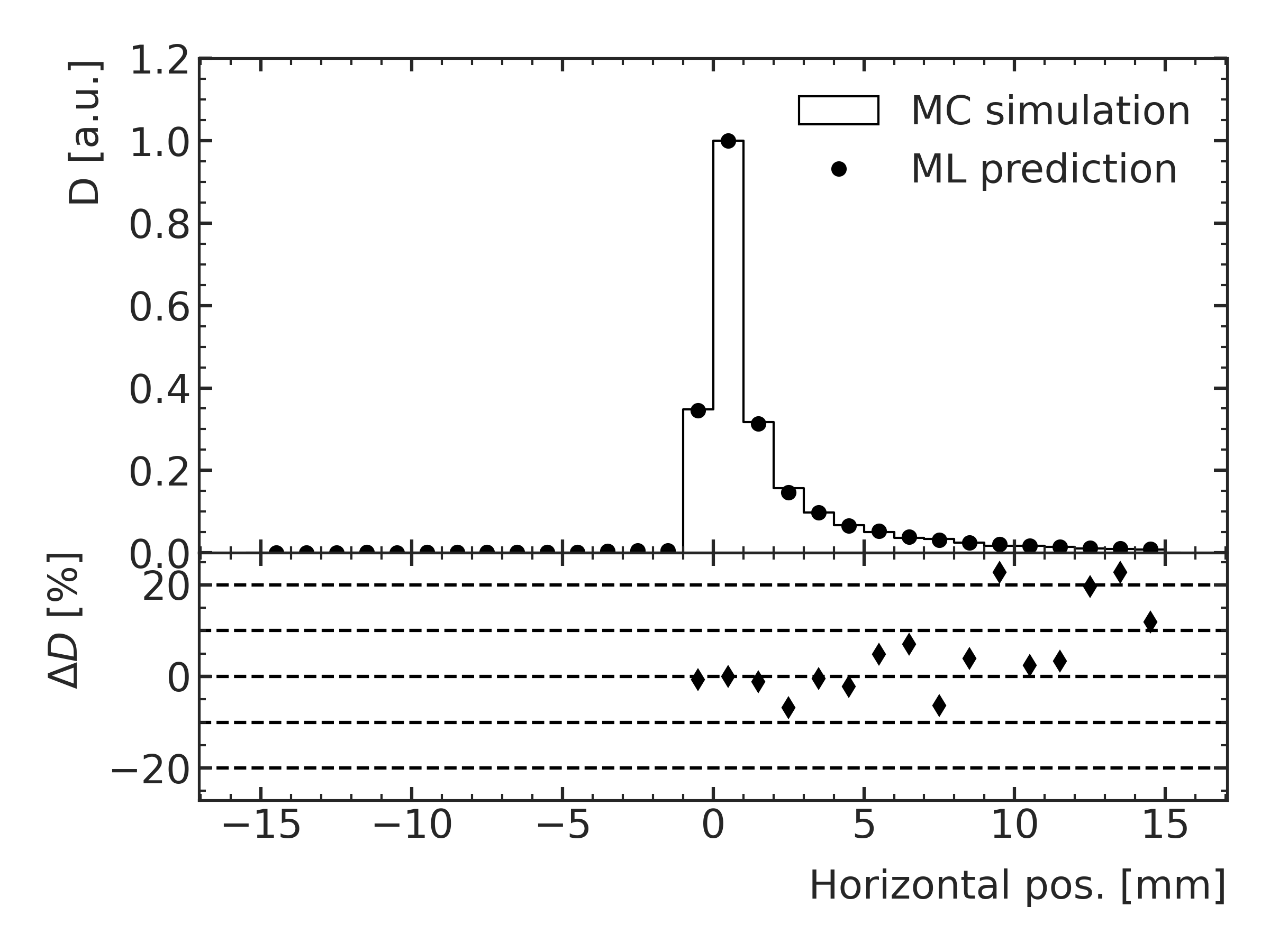}
			\caption{}
			\label{valley_predictions:e}
		\end{subfigure}
		\begin{subfigure}[t]{0.32\textwidth}
			\includegraphics[width=\linewidth]{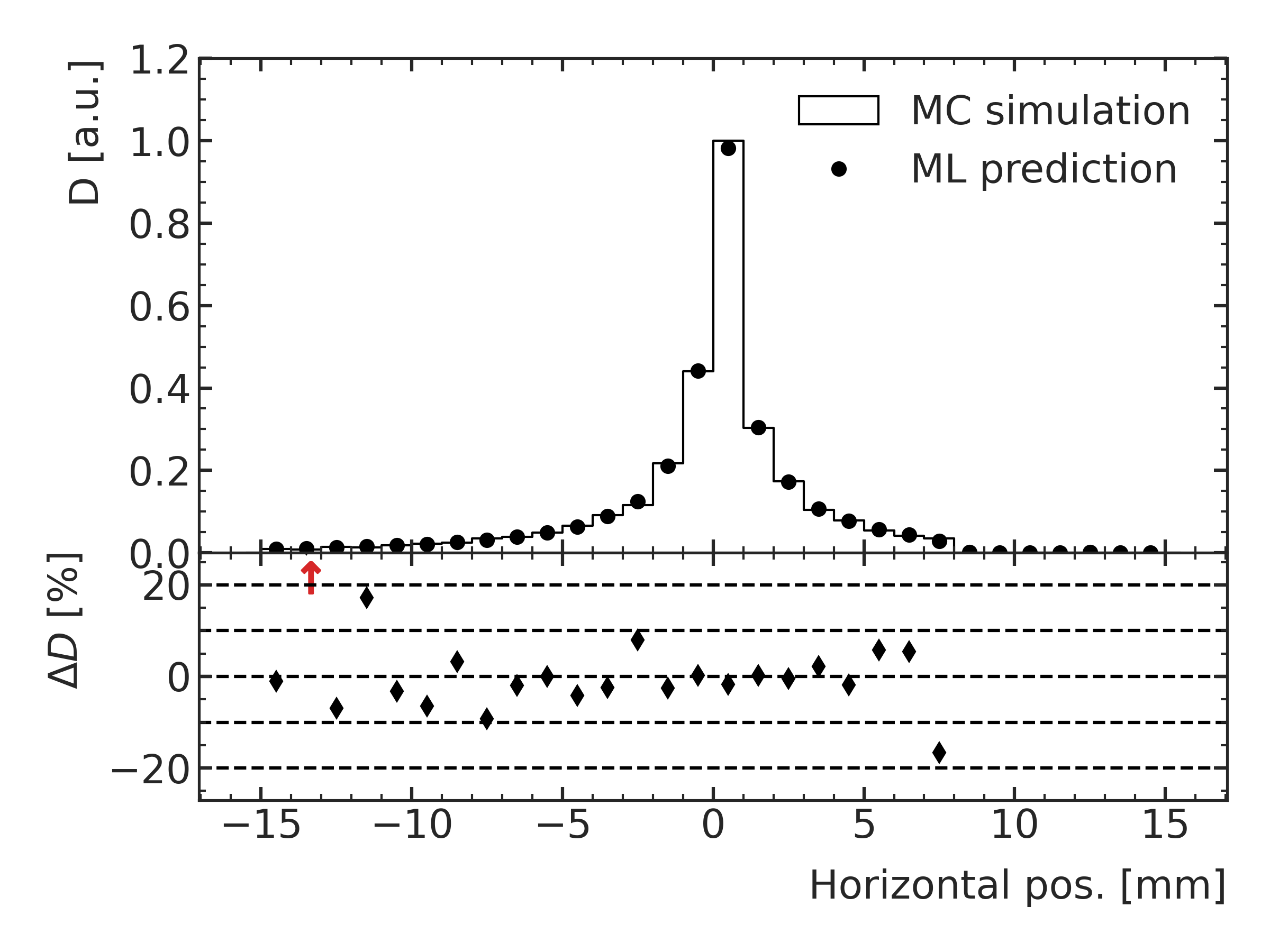}
			\caption{}
			\label{valley_predictions:f}
		\end{subfigure}
		\begin{subfigure}[t]{0.32\textwidth}
			\includegraphics[width=\linewidth]{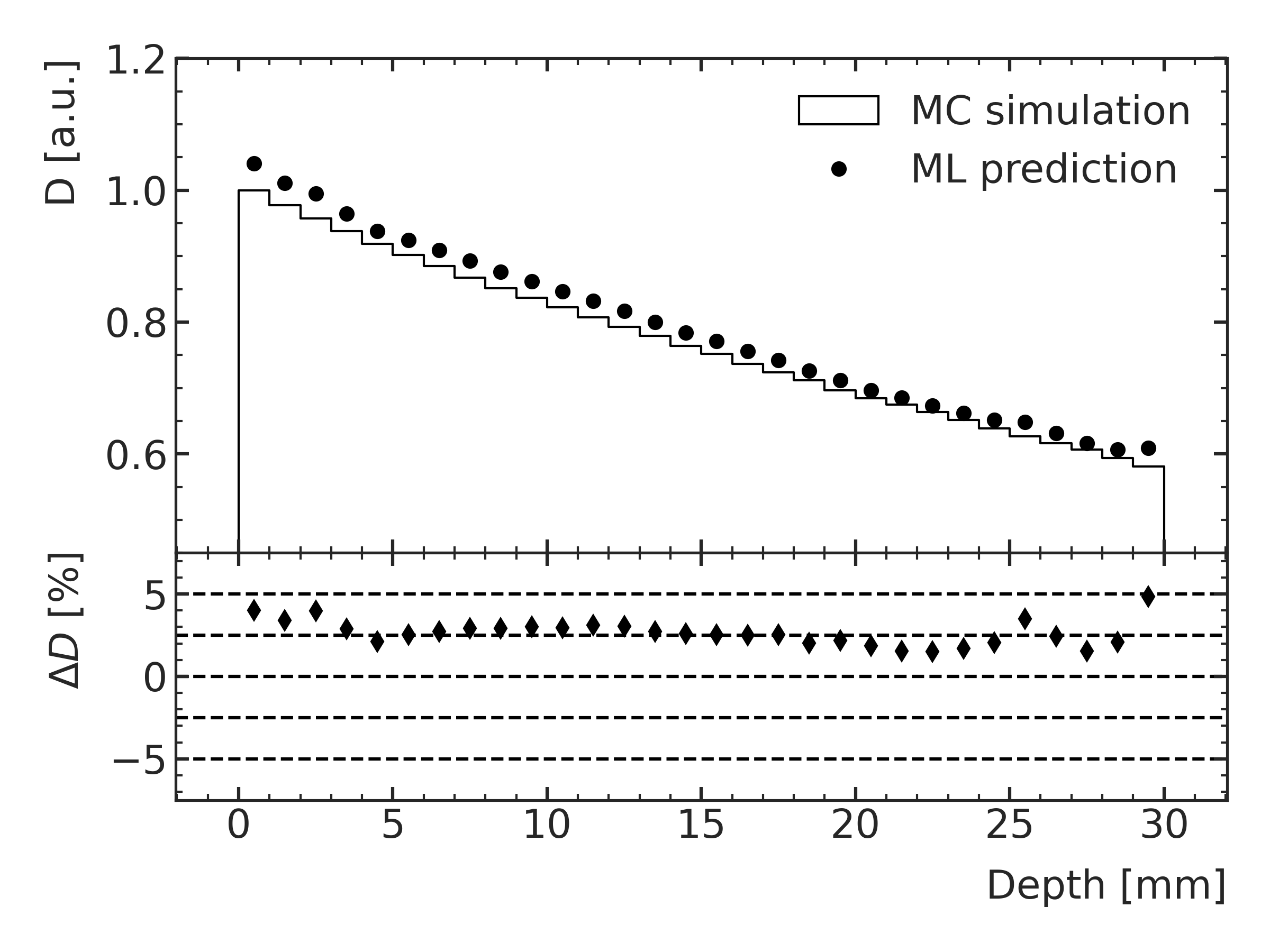}
			\caption{}
			\label{peak_predictions:a}
		\end{subfigure}
		\begin{subfigure}[t]{0.32\textwidth}
			\includegraphics[width=\linewidth]{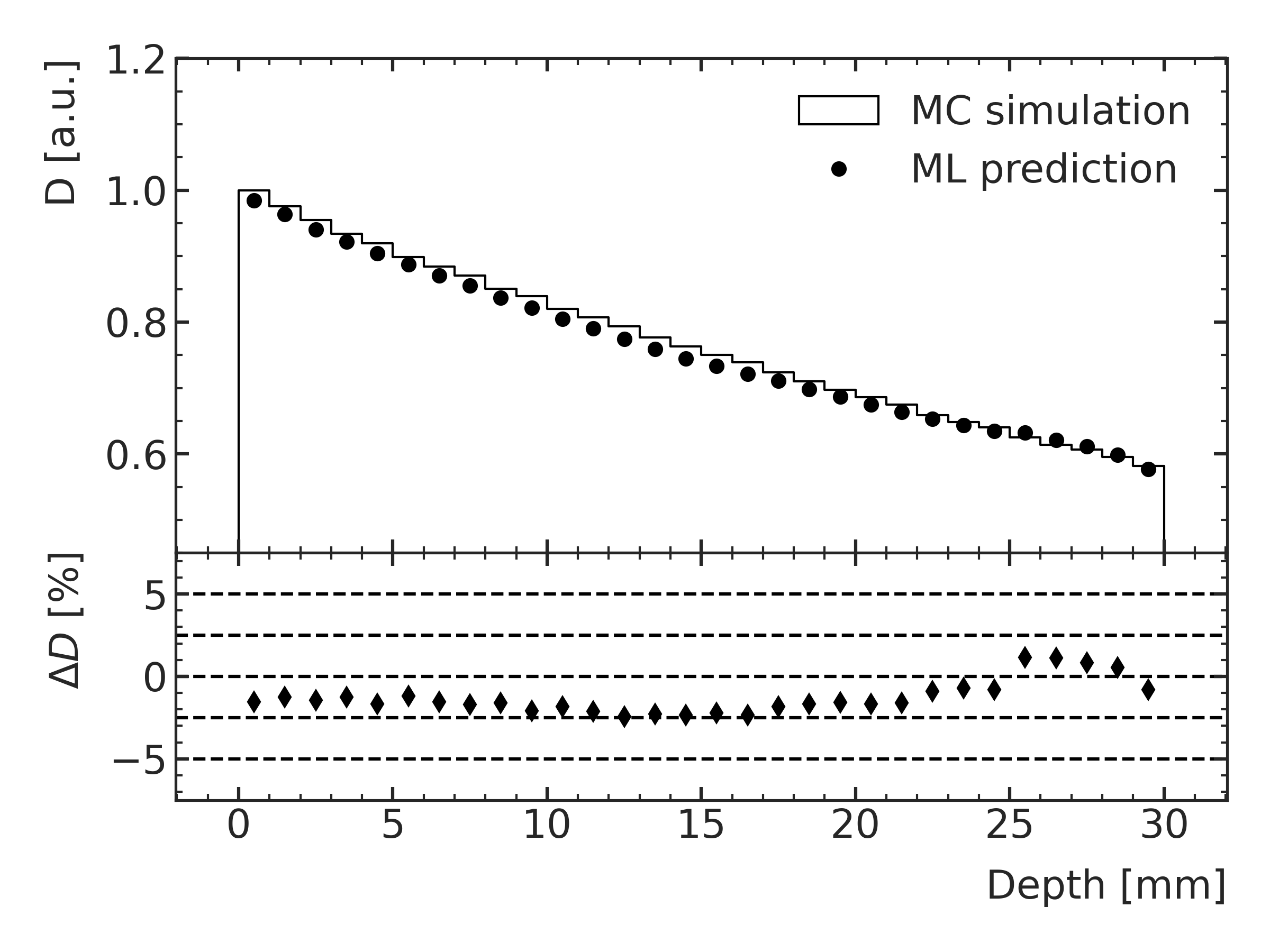}
			\caption{}
			\label{peak_predictions:b}
		\end{subfigure}
		\begin{subfigure}[t]{0.32\textwidth}
			\includegraphics[width=\linewidth]{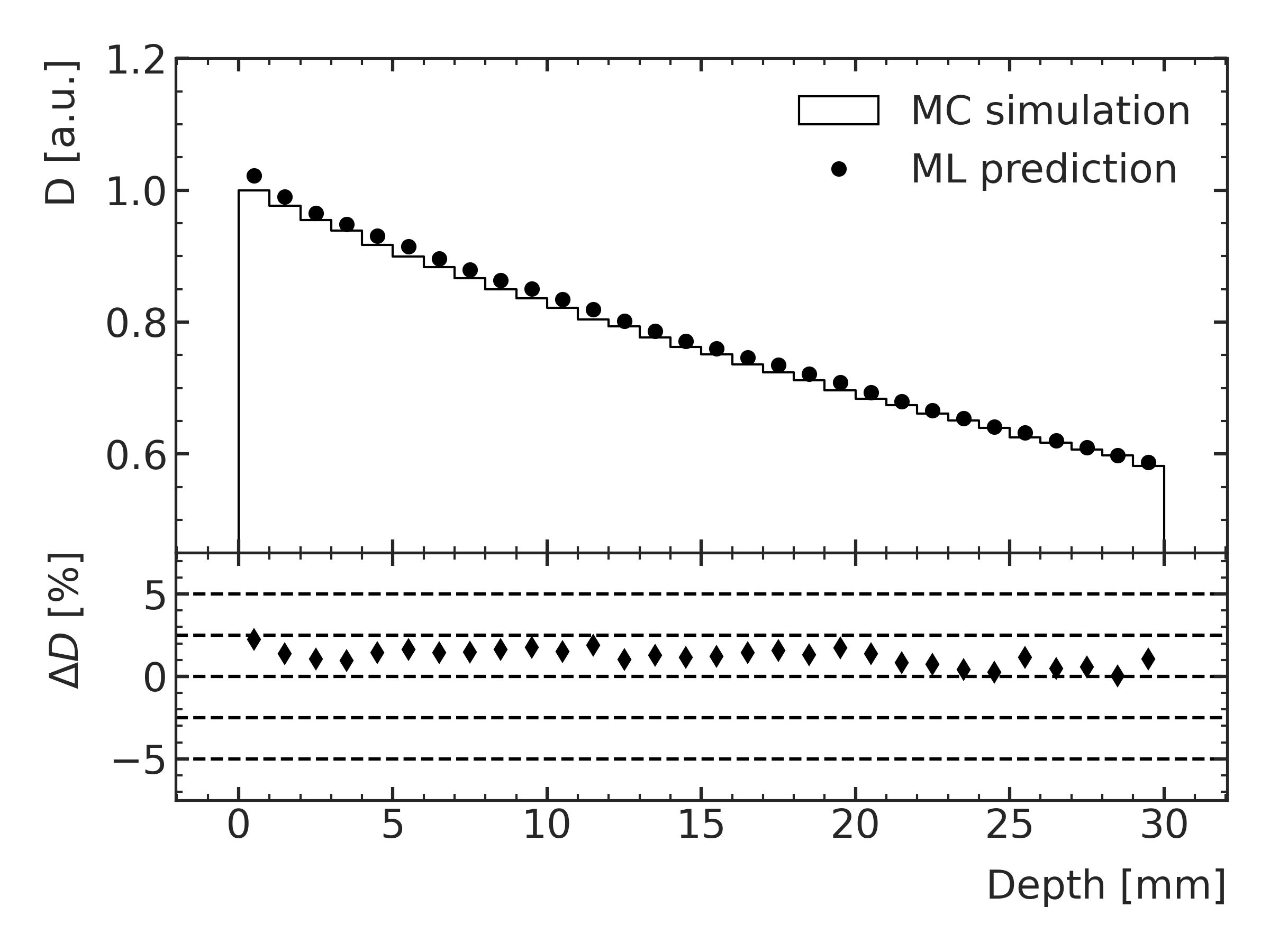}
			\caption{}
			\label{peak_predictions:c}
		\end{subfigure}
		\begin{subfigure}[t]{0.32\textwidth}
			\includegraphics[width=\linewidth]{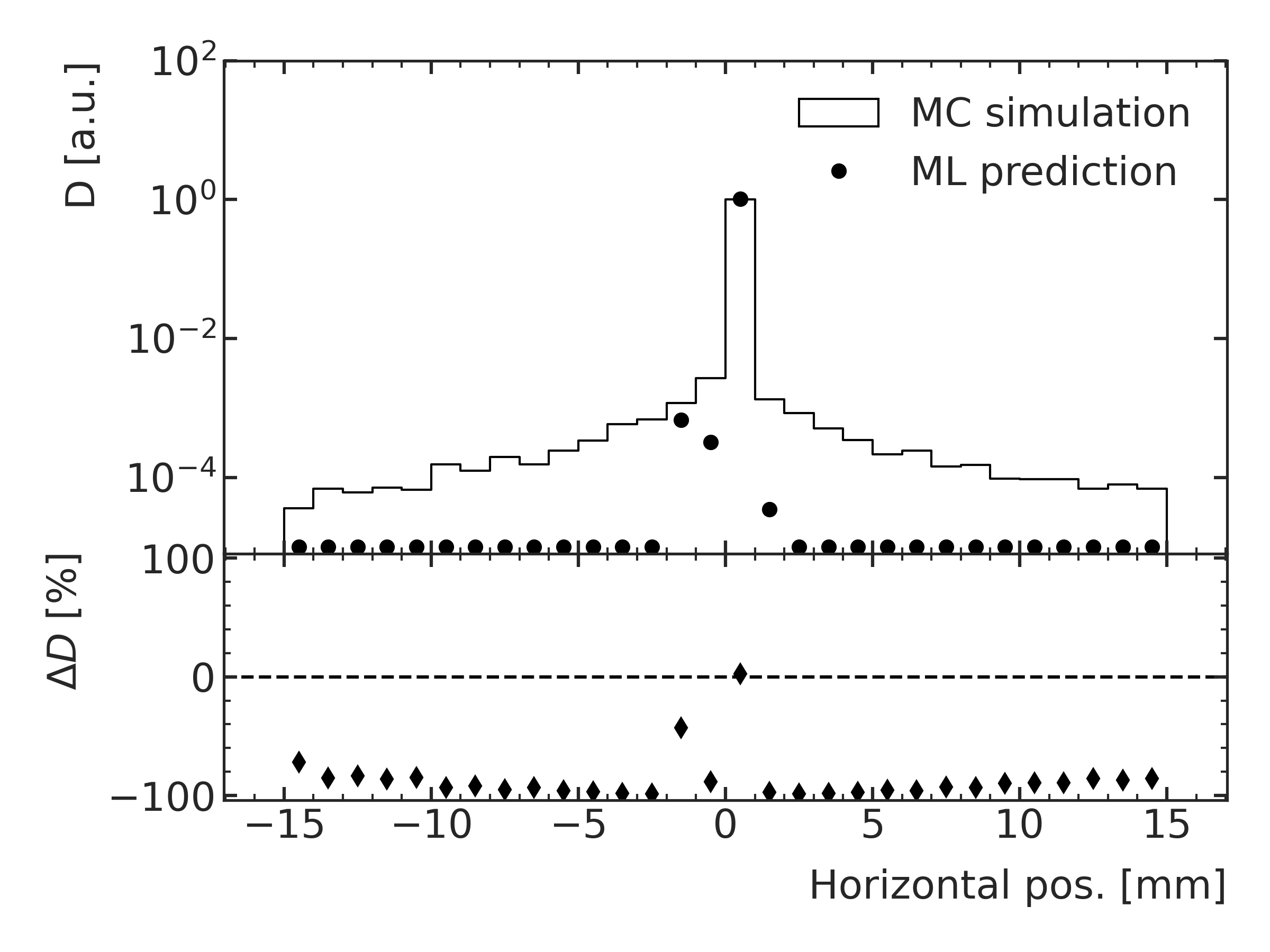}
			\caption{}
			\label{peak_predictions:d}
		\end{subfigure}
		\begin{subfigure}[t]{0.32\textwidth}
			\includegraphics[width=\linewidth]{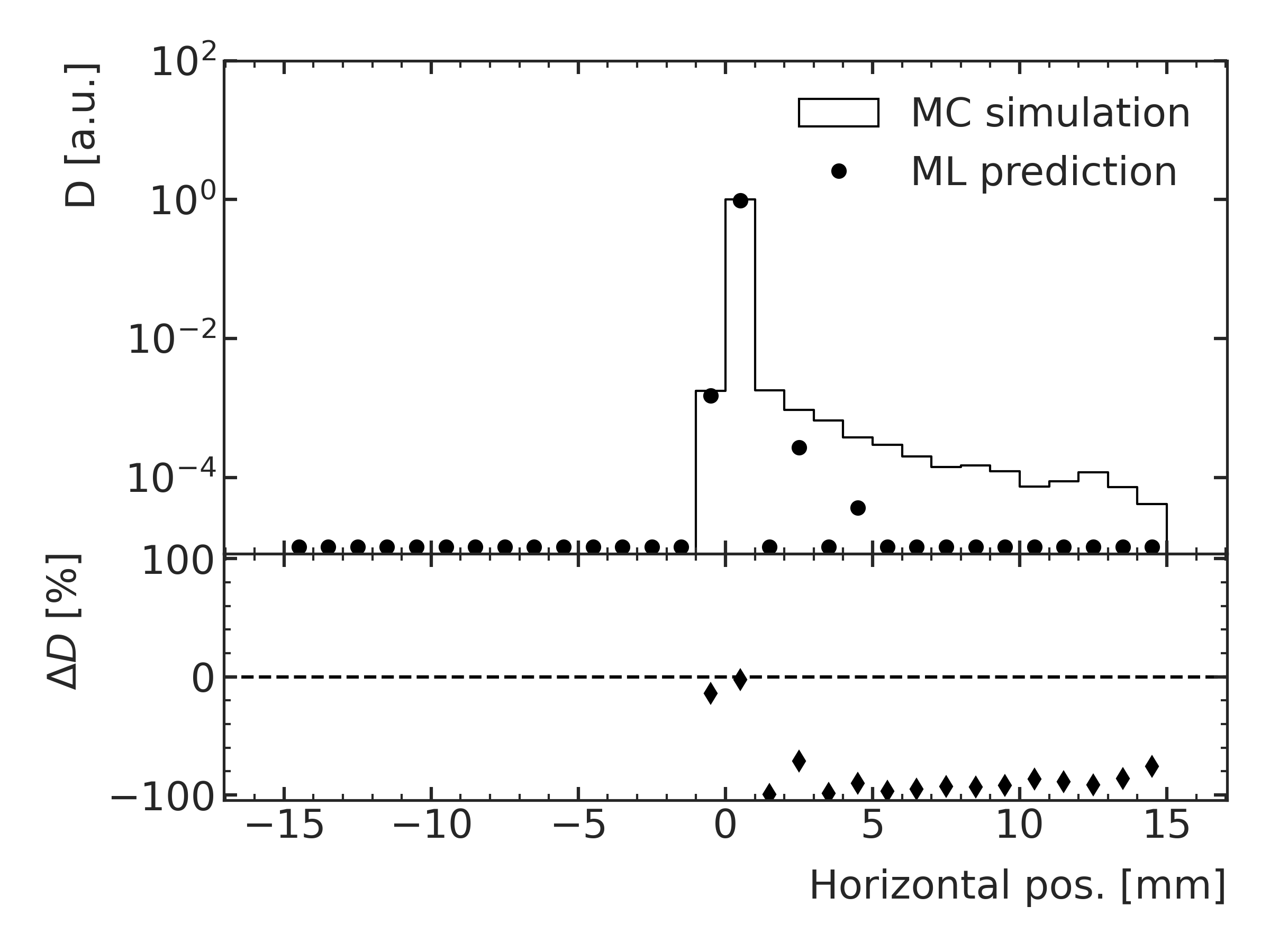}
			\caption{}
			\label{peak_predictions:e}
		\end{subfigure}
		\begin{subfigure}[t]{0.32\textwidth}
			\includegraphics[width=\linewidth]{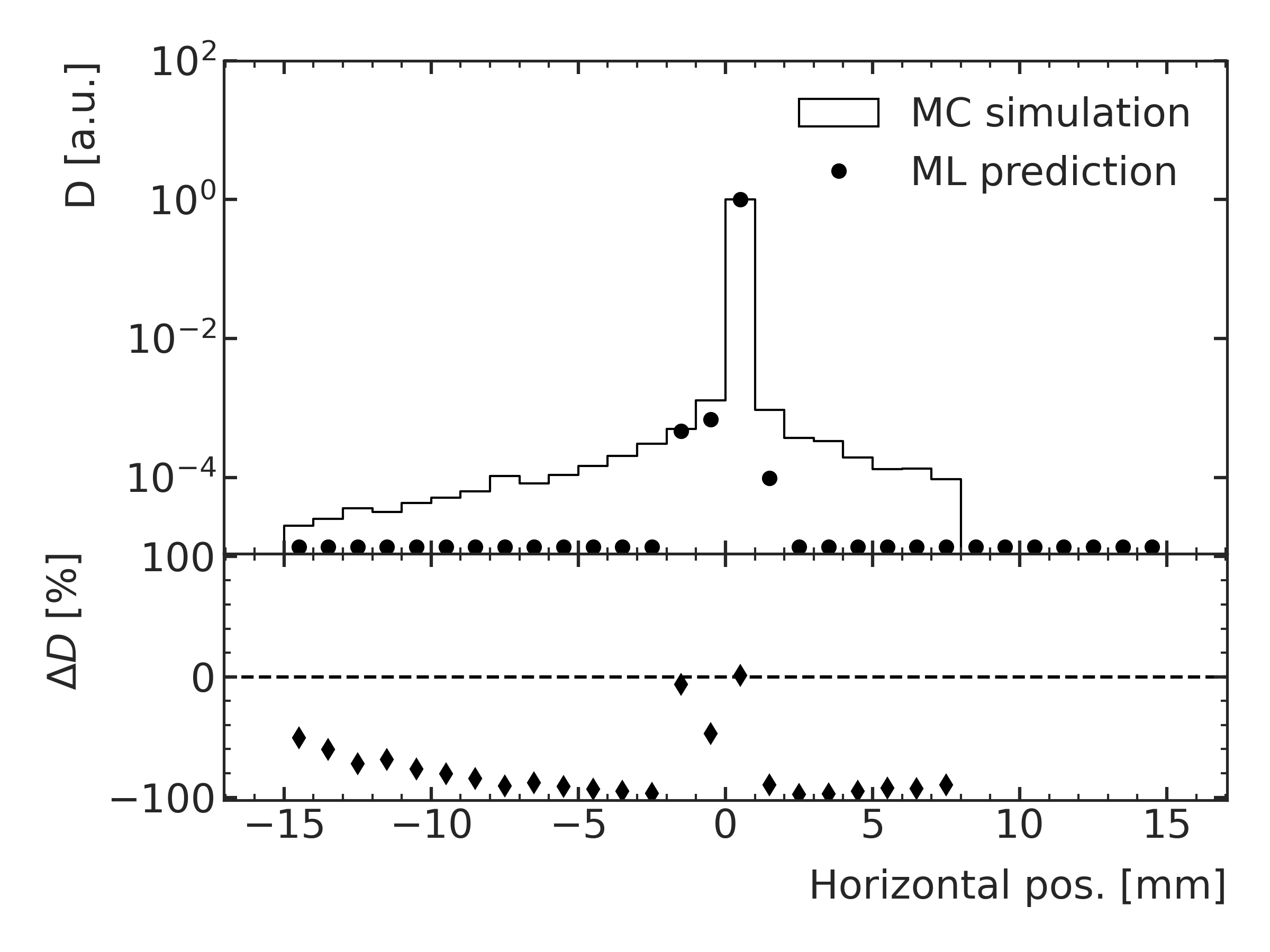}
			\caption{}
			\label{peak_predictions:f}
		\end{subfigure}
		\caption{Exemplary normalized predictions of valley (a-f) and peak (g-l) doses.  (a-c) and (g-i) show depth dose curves while (d-f) and (j-l) show lateral dose profiles at the centre of the phantom respectively. The predicted configurations are  beam translations of 55$\,\mu$m (a,d,g,j), 466$\,\mu$m (b,e,h,k), 322$\,\mu$m (c,f,i,l) for phantom translations of 0$\,$mm (a,d,g,j), -14$\,$mm (b,e,h,k), 7$\,$mm (c,f,i,l).}
		\label{valley_predictions}
	\end{figure}

	Figure (\ref{generalization_metrics}) shows a performance comparison averaged over all beam translations as a function of phantom translation: the ratio of voxels in which the ML prediction deviates less than 3\% from the MC ground truth. The values are very low due to the noisy valley data and the rather poor prediction of the peak dose network which can be inspected with exemplary predictions in Figure (\ref{valley_predictions}). Especially for the peak dose network, but also for the valley dose network, the performance on the test data set is slightly lower than for the training and validation data sets, indicating over-training. This could also be due to more complex data samples being contained in the test data.
	\\
	For further investigation, exemplary test data samples are predicted and the depth dose curves and lateral dose profiles are investigated in Figure (\ref{valley_predictions}). All three samples shown are not part of the training, but the sample resulting in the Figures (\ref{valley_predictions:c}, \ref{valley_predictions:f}, \ref{peak_predictions:c}, \ref{peak_predictions:f}) is of special interest as neither the microbeam translation ($322\,\mu$m), nor the phantom translation (7$\,$mm) are included in the training of the network. Nevertheless, the drop in dose in the lateral profile is correctly predicted at the edge of the phantom. Much of the scattering has to be attributed to the MC data noise which could not be reduced within reasonable computing time due to the large distance from the microbeams. The ML model provides a smoother prediction as it was trained on many different MC simulations allowing it to learn the average from the data. This capability, together with the notation of the extensive computing time requirements for accurate valley dose distributions emphasizes the value of such an ML model for the valley doses.
	\\
	The peak doses are predicted with a lower quality compared to the valley doses. The peak depth dose cuves exhibit deviations of up to 5\% and the lateral profile shows many entries which are clipped to $D=1\cdot 10^{-5}$ which is assigned during the prediction when the network tries to output a dose value below 0 which is not physically meaningful. 
	
	\section{Discussion\label{section:discussion}}
	
	Using the macro voxel approach, microbeam array irradiation fields can be easily constructed from single microbeams by superposition. For this, the microbeams are required to be produced in the pre-defined peak-to-peak distance between two peak scoring volumes. The differences between the presented dose distributions from a full microbeam field simulation and a superposition of multiple versions of the central microbeam are to be investigated in future studies.
	\\ 
	The 3D U-Net model was able to predict both the peak and the valley doses. Especially in the case of the peak doses though, the predictions which are further away from the central beam are subject to large deviations. As the peak doses do not influence neighbouring peak doses as much as valley doses do, as discussed in this study, this is not a major problem for the dose prediction. Nevertheless, the peak dose network should be investigated more carefully. 
	\\
	A feature of the ML model worth to note is the fast prediction. Given batch processing of currently 16 density matrices (in future CT matrices), a full irradiation field could be predicted in around 40 seconds which is significantly faster than existing methods that allow accurate dose predictions \cite{Debus2017,Donzelli2018}. 
	\\
	A proper estimate of the performance of the model given more complex training data is difficult do provide and will need more detailed investigations. So far, only a very simplified sub-space of possible microbeam applications has been investigated. The inclusion of more sophisticated models will be an important next step to showcase the full potential of this approach.
	
	\section{Conclusion\label{section:conclusion}}
	The presented macro voxel method to score peak and valley doses for microbeam radiation therapy allows for informed treatment planning without handling large dose scoring arrays which are both impractical and memory intensive. It is also possible to integrate the presented method into existing MC simulations as well. 
	\\
	Two 3D U-Nets are trained to predict the peak and valley dose distribution in a 30x30x30 voxel matrix where each voxel has an edge length of 1$\,$mm. The predictions are fast and reasonably accurate but further investigations are required towards the applicability of the presented approach for more complex microbeam radiation therapy irradiation scenarios.
	
	\section*{Acknowledgements}
	The authors gratefully acknowledge the computing time provided on the Linux HPC cluster at Technical University Dortmund (LiDO3), partially funded in the course of the Large-Scale Equipment Initiative by the German Research Foundation (DFG) as project
	271512359.
	
	\bibliography{2022_IUPESM_proceedings.bib}

\end{document}